\documentclass[
  reprint,
  superscriptaddress,
  amsmath,amssymb,
  aps,
  pre,
  floatfix
]{revtex4-2}

\usepackage{graphicx}
\usepackage{dcolumn}
\usepackage{bm}
\usepackage{hyperref}
\begin{document}

\title{Frequency bursts in adaptive delay-coupled oscillators}


\author{Yu Wang}
\affiliation{Department of Mathematics, Humboldt-Universit{\"a}t zu Berlin,
10099 Berlin, Germany}
\affiliation{Potsdam Institute for Climate Impact Research,
14473 Potsdam, Germany}

\author{Jan Sieber}
\affiliation{Department of Mathematics and Statistics, University of Exeter,
Exeter EX4 4QF, United Kingdom}

\author{Jinde Cao}
\affiliation{School of Mathematics, Southeast University,
Nanjing 210096, China}

\author{J{\"u}rgen Kurths}
\affiliation{Department of Physics, Humboldt-Universit{\"a}t zu Berlin,
10099 Berlin, Germany}
\affiliation{Potsdam Institute for Climate Impact Research,
14473 Potsdam, Germany}

\author{Serhiy Yanchuk}
\affiliation{School of Mathematical Sciences, University College Cork,
Cork T12 XF62, Ireland}
\affiliation{Potsdam Institute for Climate Impact Research,
14473 Potsdam, Germany}

\begin{abstract}
We report on frequency bursting oscillations in a system of phase oscillators
with adaptive and delayed coupling.
Adaptation of the coupling strengths is considered slow and depends on the
phase shift between the oscillators.
We find that, due to the combined effects of adaptation, collective dynamics,
and time delays, the system robustly achieves a state in which the oscillator
frequencies are nearly synchronized but detuned by an integer number of small
adaptation frequencies.
We demonstrate that this quantization of the detuning is caused by alternating
slow and fast transitions.
Moreover, the observed motions take the form of bursts of instantaneous
frequency, and the number of spikes in each burst corresponds to the
quantization level of the detuning.
We provide a fast--slow analysis of this phenomenon and explain the mechanisms
behind the emergence of bursts.
Our findings indicate that these frequency bursting oscillations are robust and
exist stably within finite parameter regions.
\end{abstract}

\maketitle

\maketitle

\section{Introduction}
Real-world dynamical networks often possess adaptive interactions, in which the node states coevolve with the coupling strengths \cite{berner2023adaptive,gross2008adaptive}. 
The node dynamics depend on the network structure, while the network structure is modified in response to the node dynamics, thereby forming a feedback loop between dynamics and connectivity. 
Such coevolution typically  occurs across physical, chemical, biological, and social systems \cite{yanchuk2025focus,Martens2017a,schweitzerSocialPercolationRevisited2021, Kuehn2019a}. 
Neuronal networks provide a prominent example for this. 
There, the spiking activity modifies the synaptic efficacy through plasticity, and the resulting connectivity in turn regulates neuronal activity and rhythms \cite{markram1997regulation,song2000competitive}. 

The interactions between nodes in coupled systems are often affected by delays.
 Signal propagation, material transport, sensing, and information processing all proceed with finite speeds \cite{erneux2009applied,yanchuk2017spatiotemporal,martins_excitability_2024}. 
As a result, the evolution of a node depends not only on the instantaneous state of the system but also on its past states, rendering delay-coupled systems effectively infinite-dimensional \cite{hale1993introduction}. 
In coupled oscillator systems, delays can shift stability boundaries, induce stability switches between phase-locked modes, and generate multiple coexisting synchronized states with delay-dependent frequencies \cite{schuster1989mutual,niebur1991collective,kim1997multistability,yeung1999time,campbell2012phase,Yanchuk2005}. 
These effects are crucial in neuronal networks, where propagation delays and synaptic plasticity jointly shape synchronization and multistability \cite{timms2014synchronization,madadiasl2018delay,madadiasl2023transitions}, as well as in semiconductor-laser networks, where optical delays control phase locking, synchronization, and dynamical mode selection \cite{soriano2013complex,flunkert2012chaos}. 
Delayed interactions are also widespread in ecological systems, where the response delay in species interactions modifies stability and generates oscillatory population dynamics \cite{pigani2022delay}.

Although adaptive coupling and interaction delay can interact to impact dynamics, their effects have mainly been investigated separately.  
Adaptive oscillator networks can exhibit splay states, synchronized clusters, and hierarchically organized multiclusters \cite{seliger2002plasticity,aoki2009coevolution,picallo2011adaptive,berner2019multiclusters,berner2019hierarchical}, to name just a few phenomena. 
Asymmetry in the adaptation rules further expands these dynamics, producing recurrent synchronization, chaotic switching between frequency clusters, and cluster bursting \cite{thiele2023asymmetric,sales2024recurrent,wei2024bursting}. 
Delayed coupling independently gives rise to coexisting phase-locked states \cite{schuster1989mutual}, reappearance of periodic orbits \cite{yanchuk2009delay}, and multistable jittering dynamics \cite{klinshov2015multistable}.
 In Stuart--Landau oscillator networks, adaptive adjustment of the coupling phase can select specific synchronized states from a multistable regime \cite{selivanov2012adaptive}.
In neuronal systems with spike-timing-dependent plasticity, transmission delays affect both synchronization patterns and connectivity structures generated by adaptation \cite{timms2014synchronization,madadiasl2018propagation,madadiasl2018delay,madadiasl2023transitions}. 
Nevertheless, delayed adaptive systems remain much less understood than their non-delayed counterparts. 
Addressing this issue requires accounting for delay-induced infinite dimensionality, the separation between fast phase dynamics and slow adaptation, and the multiplicity of phase-locked oscillations. 

Bursting oscillation is a typical phenomenon in multiscale systems.
In neuronal systems, bursts of action potentials contribute to a reliable synaptic transmission and neural information coding \cite{zeldenrust2018neural}.
In semiconductor lasers, a delayed optical feedback can generate low-frequency fluctuations and regular pulse packages in which slower envelope dynamics organize fast intensity oscillations \cite{heil2001dynamics,ruschel2017chaotic,niiyama2022power}.
Such bursting oscillations have also been observed in electrochemical systems \cite{organ2003bursting,kiss2006electrochemical}.
Despite their different physical origins, these phenomena share a characteristic separation of timescales \cite{desroches2022classification}.
In phase dynamics, abrupt phase slips can likewise appear as localized spikes in the instantaneous frequency \cite{hurtado2004statistical}.
This suggests a frequency-domain form of bursting in which the slow evolution organizes repeated fast transitions into groups of instantaneous frequency spikes.

In this work, we explore the combined influence of adaptation and delay. We find a phenomenon in which two interacting oscillators stay synchronized with each other over long times apart from rare phase slips, resulting in a small detuning of their mean frequencies $\Omega_1$ and $\Omega_2$, proportional to the inverse of the adaptation timescale,  $\Omega_1 - \Omega_2 \sim \varepsilon\ll 1$.
This detuning is not unique and can be quantized as $m\varepsilon$, where $m=1,2,\cdots$. 
Furthermore, the oscillators' instantaneous frequencies exhibit bursting behavior, with the number of bursts corresponding to the quantization level $m$. 
Using multiscale methods, we describe the geometric mechanisms behind the emergence of frequency bursts and near-synchronous, quantized detuning. These mechanisms involve slow drifts along quasistationary states (slow manifolds of relative equilibria) and fast transitions between these states. The interplay between the adaptation timescale $1/\varepsilon$ and the time delay $\tau$ leads to the emergence of bursts with a particular number of spikes. For brevity, we will call the corresponding oscillations FB (frequency bursting) or FB oscillations.

We consider a reduced variant of the adaptive Kuramoto--Sakaguchi model, which describes the coevolution of fast oscillator phases and slowly adapting couplings, while retaining a sufficiently simple structure for analytical investigations 
\cite{kuramoto1984chemical,sakaguchi1986soluble,aoki2009coevolution,berner2023adaptive,wolfrum_turbulence_2016,thiele2023asymmetric,sawicki2023perspectives,cestnikContinuumLimitAdaptive2025,sharmaSynchronizationTransitionsAdaptive2024,Andreev2022}. 
The model has been widely used to study synchronization,  multistability, and adaptive cluster formation. Introducing interaction delay into this framework, thus provides a paradigmatic system for investigating adaptation-, and delay-related  bursting dynamics. 
The slow-fast structure of the system enables us to employ multiscale methods \cite{kuehn2015multiple,wechselberger2020geometric}. 

This work is organized as follows. 
Section~\ref{sec:model} introduces the model of two adaptively delay-coupled phase oscillators. 
Section~\ref{sec:relative_phase_locking} introduces FB oscillations and classifies them according to the integer relative  winding jumps over one slow period. 
Section~\ref{sec:RE_CMs} derives the critical manifolds of the fast relative equilibria and identifies the relative equilibria of the full slow-fast system. Section~\ref{sec:RPO_Phase_Locking} discusses how the FB oscillations arise from a slow drift along stable critical manifold sheets and fast  jumps. 
Section~\ref{sec:parameter_organization_RPOs} performs a continuation of stable FB oscillation families and investigates their multistability. 
Finally, in Sec.~\ref{sec:conclusions}, we summarize and discuss our results.

\section{Adaptive delay-coupled phase oscillator model}\label{sec:model}
We consider the following paradigmatic model of two adaptively delay-coupled phase oscillators:
\begin{align}
\label{eq:phi1}
\dot{\phi}_{1}(t)&=\omega_{1}-\kappa_{1}\sin\left(\phi_{1}(t)-\phi_{2}(t-\tau)+\alpha\right),\\
\label{eq:phi2}
\dot{\phi}_{2}(t)&=\omega_{2}-\kappa_{2}\sin\left(\phi_{2}(t)-\phi_{1}(t-\tau)+\alpha\right),\\
\label{eq:kappa1}
\dot{\kappa}_{1}(t)&=-\epsilon\left[\kappa_{1}-a_1\sin\left(\phi_{1}(t)-\phi_{2}(t)+\beta_1\right)\right],\\
\label{eq:kappa2}
\dot{\kappa}_{2}(t)&=-\epsilon\left[\kappa_{2}-a_2\sin\left(\phi_{2}(t)-\phi_{1}(t)+\beta_2\right)\right],
\end{align}
where $0<\epsilon\ll1$  is a small parameter that expresses the difference in the timescales between the dynamics of the fast phases $\phi_1,\phi_2$, and the evolution of the slow coupling strengths $\kappa_1,\kappa_2$.
$\omega_i$ is the natural frequency of the $i$-th oscillator, while $\alpha$ measures the fixed phase shift in the coupling between the oscillators and $\tau$ is the delay in the coupling. 
The parameters $a_1$, $a_2$ determine the amplitudes, and  $\beta_1$, $\beta_2$ are the phase shifts in the adaptation rules. 
In what follows, we consider  the case $\beta_1=0$, and denote $\beta_2=-\pi/2$. 
Therefore, the two adaptation laws are different, $\kappa_1$ follows a causal rule, whereas $\kappa_2$ follows a Hebbian-like rule
\cite{aoki2009coevolution,aoki2011self,berner2019multiclusters}.
The parameters $\omega_1$ and $\omega_2$ can be rescaled to 0 and 1, respectively, by non-dimensionalization, see \ref{sec:nondim} for the details.

System \eqref{eq:phi1}--\eqref{eq:kappa2} provides a minimal model for coupled oscillators with delayed coupling, adaptive coupling strengths and different adaptation rules. The model has a slow-fast structure with the fast variables $(\phi_1,\phi_2)$ and the slow variables $(\kappa_1,\kappa_2)$. 
Although referred to as phase oscillators, the variables $\phi_i$ are actually rotators that can rotate around a circle.
The fast subsystem (layer system) is infinite-dimensional due to the presence of the time-delay $\tau$ in the coupling term, while the adaptation rule is governed by  ordinary differential equations for $\kappa_i$. 
Such infinite-dimensional fast dynamics are common in models of semiconductor lasers with optical feedback \cite{lang1980external,yanchuk2010multiple,bauer_nonlinear_2004,soriano2013complex}.

System \eqref{eq:phi1}--\eqref{eq:kappa2} has the phase-shift symmetry $\phi_i\mapsto\phi_i+\psi~(\psi\in \mathbb{R}/2\pi\mathbb{Z})$. 
This symmetry gives rise to solutions of the form $\phi_i(t)=\Omega t + \xi_i(t)$, $\kappa_i(t)$. Such solutions are called relative equilibria when the $\xi_i$ and $\kappa_i$ are constant, and relative periodic solutions when these functions are periodic \cite{krupa1990bifurcations,lamb2007normal,yanchuk2013relative}.

\section{The phenomenon: Frequency bursting (FB)}
\label{sec:relative_phase_locking}

\begin{figure}
    \centering
\includegraphics[width=0.9\linewidth]{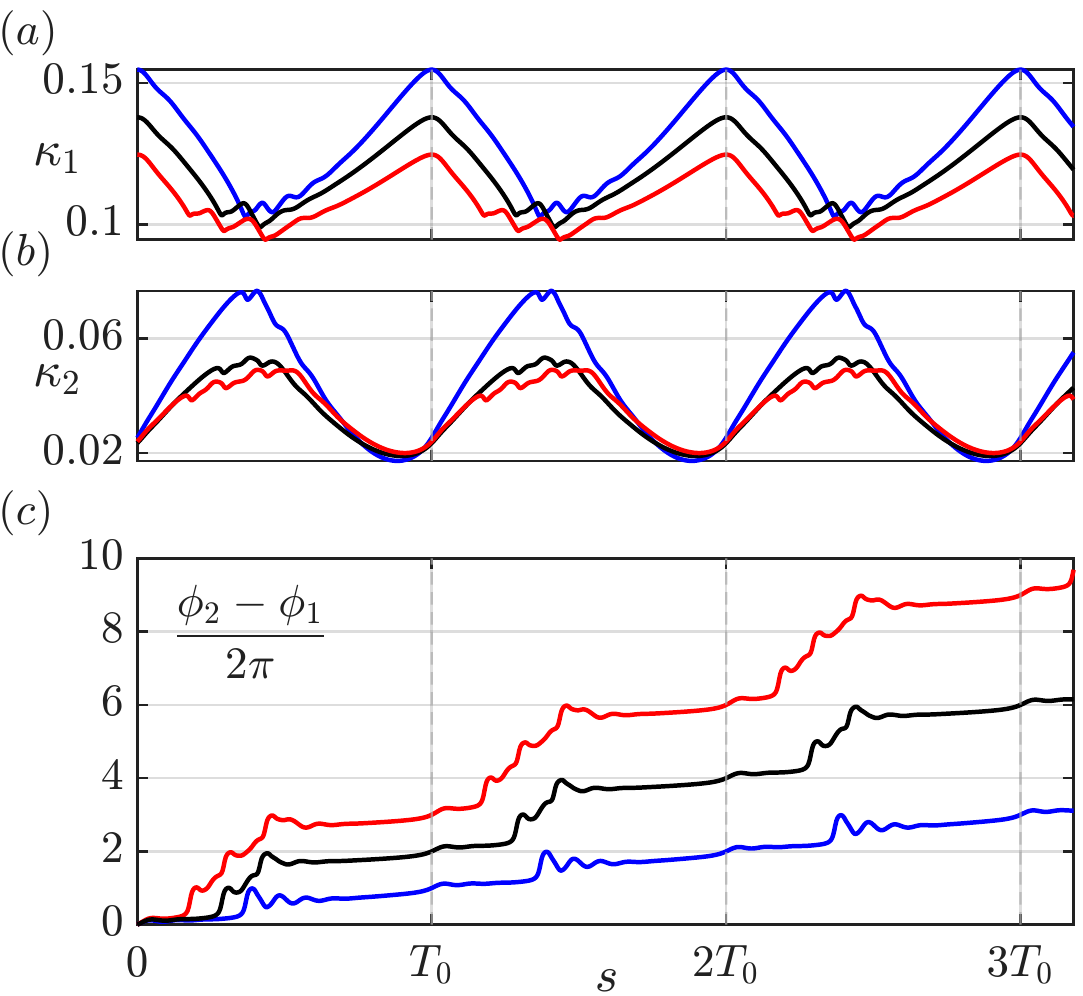}
\includegraphics[width=0.9\linewidth]{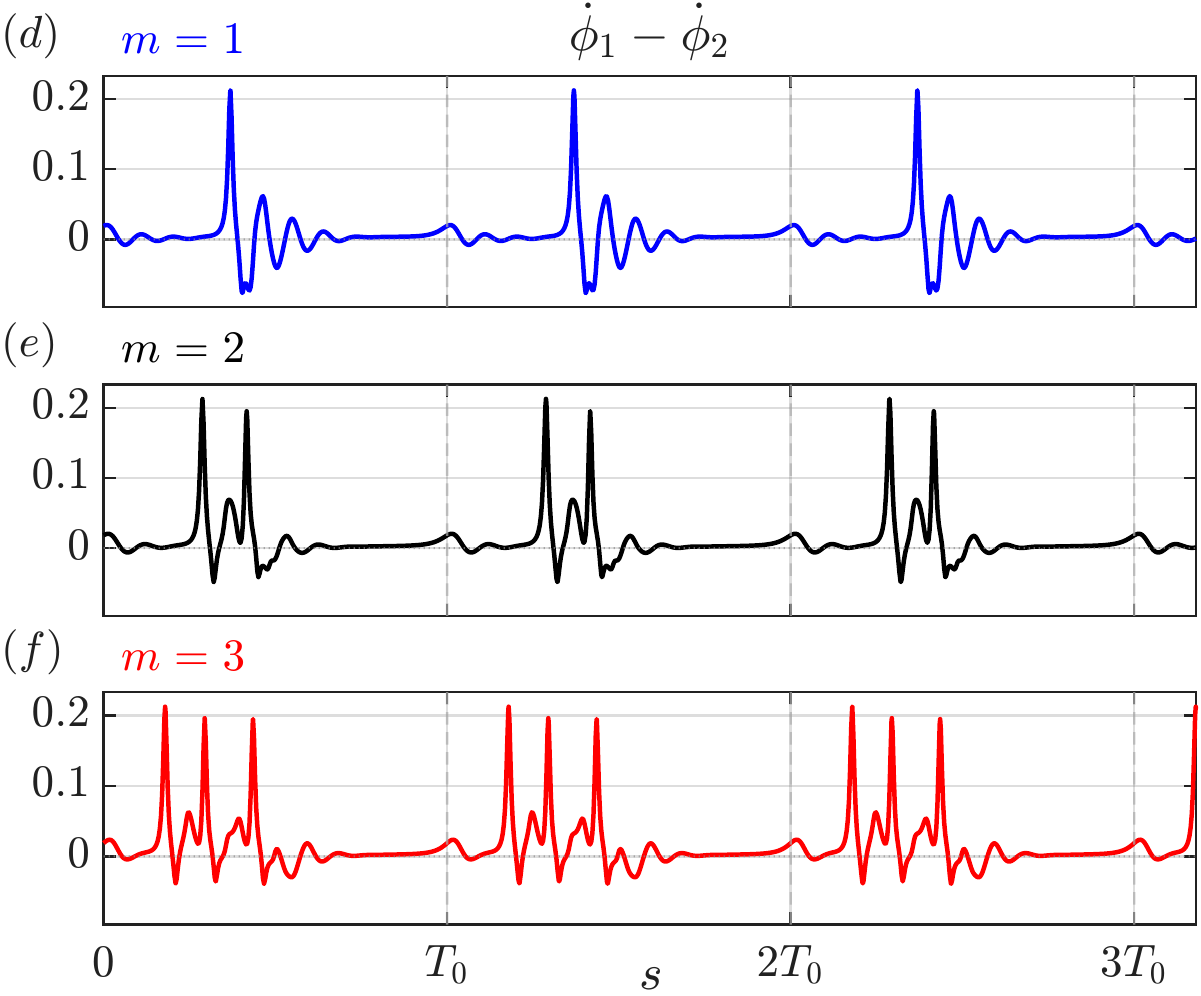}
    \caption{
     Representative frequency-bursting oscillations with winding numbers $m=1$ (blue), $m=2$ (black), and $m=3$ (red).
     For comparison, time is normalized as $s=T_0(t-t_0)/T_m$, where $T_m$ is the period of each oscillation
     and $T_0$ is the common normalized period.
     Panels (a) and (b) show the adaptive coupling variables $\kappa_1$ and $\kappa_2$, while panel (c) shows the accumulated relative winding $W_\Delta(t)$ defined in Eq.~\eqref{eq:WindingJump}, with a net increase of $m$ over each slow period.
     Panels (d)--(f) show the corresponding instantaneous relative frequencies $\dot{\phi}_1(t)-\dot{\phi}_2(t)$ for $m=1,2,3$, respectively.
     The three cases are $m=1$ for $(a_1,a_2)=(0.4,0.4)$, $m=2$ for $(a_1,a_2)=(0.3,0.27)$, and $m=3$ for $(a_1,a_2)=(0.245,0.27)$.
     The remaining parameters are $\omega_1=0.2$, $\omega_2=0.1$, $\alpha=\pi/4$, $\beta_1=0$, $\beta_2=-\pi/2$, $\epsilon=4\times10^{-4}$, and $\tau=40$.
    }
\label{fig:relative_phase_locking}
\end{figure}

This section introduces the main phenomenon of frequency bursting (FB), see~Fig.~\ref{fig:relative_phase_locking}, in which two interacting oscillators stay synchronized with each other over long times of order $1/ \epsilon$ apart from rare and fast phase slips.

Periodic FB oscillations of system \eqref{eq:phi1}--\eqref{eq:kappa2} have the form 
\begin{align}
&\phi_1(t) = \Omega t+\theta_1(t),\quad \theta_1(t)=\theta_1(t+T_m), \label{eq:rel_phi1} \\
&\phi_2(t) = \Omega t+\frac{2\pi m}{T_m}t+\theta_2(t),\quad \theta_2(t)=\theta_2(t+T_m), \label{eq:rel_phi2}\\
&\kappa_i(t) = \kappa_i(t+T_m),\qquad i=1,2;~m=1,2,\cdots
\label{eq:rel_kappa_i}
\end{align}
where the modulations $\theta_i(t)$ have specific slow-fast properties described below. 
These solutions have the modulation period $T_m$, which will be shown to be proportional to $1/\varepsilon$. 
We call this `modulation period', since this becomes a truly periodic solution in the reference frame, where the phases oscillate with the frequency $\Omega$, i.e., $\phi_i(t)-\Omega t$. 
The slow coupling weights $\kappa_{1,2}(t)$ are exactly $T_m$-periodic (Figs.~\ref{fig:relative_phase_locking}(a) and (b)), while the fast phases possess mean frequencies (phase velocities)  $\Omega_1=\Omega$ and $\Omega_2=\Omega + 2 \pi m/T_m$, respectively, and they are $T_m$-periodically modulated by $\theta_1(t)$ and $\theta_2(t)$. 
Here, we define the mean frequencies as $\Omega_i=\lim_{T_m\to\infty}\{[\phi_i(t+T_m)-\phi_i(t)]/ T_m\}$.

The approximately antiphase modulation of $\kappa_1$ and $\kappa_2$ in Figs.~1(a) and (b) results from the different adaptation rules with $\beta_1=0$ and $\beta_2=-\pi/2$.
The phase difference $\phi_2-\phi_1$ alternates between slow drifts and fast transitions (Fig.~\ref{fig:relative_phase_locking}(c)). 
The flat parts correspond to the slow drift, during which the relative phase remains approximately locked. The step-like increases correspond to fast transitions between the phase-locked states.
The slow drift and fast transitions over one period $T_m$ produce the integer increment of the winding number 
\begin{align}
W_{\Delta}(T_m)=m, \quad \text{where}~W_{\Delta}(t)=W_{\phi_2}(t)-W_{\phi_1}(t),
\label{eq:WindingJump}
\end{align}
and the winding numbers of each phase are defined by
\begin{align}
W_{\phi_i}(t)=\frac{\phi_i(t)-\phi_i(0)}{2\pi},
\quad i=1,2.
\label{eq:WindingNumber}
\end{align}

Moreover, the phase jumps correspond to the spikes in the instantaneous relative frequencies
$\dot{\phi}_1-\dot{\phi}_2$, see Figs.~\ref{fig:relative_phase_locking}(d)--(f).
During the phase-locked segments, frequencies remain close to constant or oscillate below a threshold, while the localized spikes (or bursts when $m>1$) appear during the fast transitions underlying the increments of $m$.

The phase locking in non-symmetric systems is commonly characterized by a rational ratio, $\Omega_1/\Omega_2$, of the mean frequencies, such that the oscillators complete integer numbers of oscillations over a common time interval \cite{ermentrout1981phase,ermentrout1984frequency,izhikevich2004weakly,ren2000phase}. 
For the FB oscillations considered here, the relative phase advances by an integer multiple of $2\pi$ over the period $T_m$, i.e., $\phi_2(t+T_m)-\phi_1(t+T_m)=\phi_2(t)-\phi_1(t)+2\pi m~(m\in\mathbb{Z})$.
Consequently, the ratio of the individual mean frequencies needs not be rational. Instead, their frequency difference is locked to an integer multiple of the slow adaptation frequency $2\pi/T_m$.
We characterize these FB oscillations by the integer $m$.
In our results, different values of $m$ distinguish distinct relative periodic solution families that persist and may coexist with different initial conditions.
 
This behavior can also be described as near-synchrony with a quantized mean-frequency detuning, since the mean frequencies are close $\Omega_1-\Omega_2 =2\pi m/T_m \sim m\varepsilon$, with their differences being quantized my $m$, where $m$ is the number of spikes in the burst in one period. In the example from Fig.~\ref{fig:relative_phase_locking}, $\Omega_1-\Omega_2 = 0.0052$, $0.0103$, and $0.0146$ for $\varepsilon=0.0004$. 
If we consider the non-dimensional time $(\omega_2-\omega_1)t$ introduced in Appendix~\ref{sec:nondim}, both $\Omega_1-\Omega_2$ and $\varepsilon$ need to be multiplied by  $10$.

\section{Relative equilibria and critical manifolds}\label{sec:RE_CMs}

The backbone of the fast dynamics is organized by the relative equilibria $ \phi_i(t) = \Omega t + \theta_i^*$ of the fast subsystem \eqref{eq:phi1}--\eqref{eq:phi2}. 
The families of the relative equilibria parametrized by $\kappa_1$ and $\kappa_2$ comprise the critical manifolds of the relative equilibria \cite{kuehn2015multiple}. 
The relative equilibria of the full slow-fast system are then identified within these manifolds by additionally imposing the equilibrium conditions of the slow adaptation equations.

\subsection{Critical manifolds of fast relative equilibria}\label{subsec:CM_REs}

The relative equilibria of the fast subsystem have the form
\begin{align}
\label{eq:RE_phi1}
\phi_1(t)=\Omega t+\theta_1^*,\\
\label{eq:RE_phi2}
\phi_2(t)=\Omega t+\theta_2^*.
\end{align}
Due to phase-shift symmetry, both of the values, $\theta_1^*$ and $\theta_2^*$, can be shifted simultaneously by an arbitrary amount. Therefore, only the phase difference, defined as $\theta = \theta_1^* - \theta_2^*$, needs to be found. We recall that the slow variables $\kappa_1$ and $\kappa_2$ are regarded as parameters in the fast subsystem \eqref{eq:phi1}--\eqref{eq:phi2} for $\varepsilon=0$. Substituting  \eqref{eq:RE_phi1} and \eqref{eq:RE_phi2} into \eqref{eq:phi1} and \eqref{eq:phi2} yields a system of equations for the unknown quantities $\theta$ and $\Omega$.
\begin{align}
\Omega
&=
\omega_{1}
-
\kappa_{1}\sin(\theta+\Omega\tau+\alpha),
\label{eq:Fast_phi1}
\\
\Omega
&=
\omega_{2}
-
\kappa_{2}\sin(-\theta+\Omega\tau+\alpha).
\label{eq:Fast_phi2}
\end{align}
 Then, we  solve  Eqs.~\eqref{eq:Fast_phi1}--\eqref{eq:Fast_phi2} for $\sin\theta$ and $\cos\theta$
\begin{align}
   \label{eq:sin_theta}
    &\sin\theta=-\frac{1}{{2\cos(\Omega\tau+\alpha)}}
    \left(\frac{\Omega-\omega_1}{\kappa_1}-\frac{\Omega-\omega_2}{\kappa_2}\right),\\
   \label{eq:cos_theta}
    &\cos\theta=-\frac{1}{{2\sin(\Omega\tau+\alpha)}}
    \left(\frac{\Omega-\omega_1}{\kappa_1}+\frac{\Omega-\omega_2}{\kappa_2}\right).
\end{align}
Using the relation $\sin^2\theta+\cos^2\theta =1$, we obtain the following scalar equation for the unknown frequencies  $\Omega$  of the fast relative equilibria:
\begin{equation}
\label{eq:G_gamma_new}
\begin{aligned}
    0 =& G(\Omega,\kappa_1,\kappa_2)\\=&\frac{1}{{\sin^2(\Omega\tau+\alpha)}}\left(\frac{\Omega-\omega_{1}}{\kappa_{1}}+\frac{\Omega-\omega_{2}}{\kappa_{2}}\right)^{2}\\&+\frac{1}{\cos^2(\Omega\tau+\alpha)}\left(\frac{\Omega-\omega_{1}}{\kappa_{1}}-\frac{\Omega-\omega_{2}}{\kappa_{2}}\right)^{2}-4.
\end{aligned}
\end{equation}
This representation requires $\kappa_1\kappa_2\ne0$. The cases $\kappa_1=0$ or $\kappa_2=0$ have to be treated directly from Eqs.~\eqref{eq:Fast_phi1}--\eqref{eq:Fast_phi2}.

For each parameter pair $(\kappa_1,\kappa_2)$, Eq.~\eqref{eq:G_gamma_new} admits a finite number $N_\Omega(\kappa_1,\kappa_2)$ of distinct real roots $\Omega_\ell(\kappa_1,\kappa_2)$, $\ell=1,\cdots,N_\Omega$. 
Substituting each $\Omega_\ell$ into Eqs.~\eqref{eq:sin_theta} and \eqref{eq:cos_theta} yields the corresponding phase difference $\theta_\ell(\kappa_1,\kappa_2)$.
Away from branch fold points, the relative equilibria are organized locally into critical manifold sheets. 
Fixing $\theta_1^*=0$, so that $\theta_2^*=-\theta$, we write 
\begin{align*}
C_\ell
&=
\left\{
\begin{bmatrix}
    \phi_1((\cdot);\kappa_1,\kappa_2)\\
    \phi_2((\cdot);\kappa_1,\kappa_2)
\end{bmatrix}
:
\begin{bmatrix}
    \phi_1(t;\kappa_1,\kappa_2)\\
    \phi_2(t;\kappa_1,\kappa_2)
\end{bmatrix},
~
(\kappa_1,\kappa_2)\in D_\ell
\right\}
\\
&=
\left\{
\begin{bmatrix}
    \Omega_\ell(\kappa_1,\kappa_2)t\\
    \Omega_\ell(\kappa_1,\kappa_2)t
    -\theta_\ell(\kappa_1,\kappa_2)
\end{bmatrix},
~
(\kappa_1,\kappa_2)\in D_\ell
\right\}.
\end{align*}
Here $D_\ell\subset(\mathbb{R}\setminus\{0\})^2$ denotes a connected parameter domain on which the root branch $\Omega_\ell(\kappa_1,\kappa_2)$ and the associated phase difference $\theta_\ell(\kappa_1,\kappa_2)$ are defined smoothly. Each $C_\ell$ represents a critical manifold sheet associated with the fast relative equilibrium $(\Omega_\ell,\theta_\ell)$. 
Since each sheet is parameterized by the two slow variables $(\kappa_1,\kappa_2)$, $C_\ell$ forms a collection planar sheets, with $\Omega_\ell$ and $\theta_\ell$ determined by $(\kappa_1,\kappa_2)$ along the corresponding sheet.
Note that, due to the time-delay, these manifolds $C_\ell$ are embedded in an infinite-dimensional functional space \cite{hale1993introduction}, but we do not include further theoretical aspects, which are not essential for our analysis. This infinite-dimensionality of the phase space leads to the fact that the number of sheets $N_\Omega$ of the critical manifolds grows as $\tau$ or $\kappa_j$ increase \cite{yanchuk2009delay,yanchuk2013relative}. 

Another consequence of the time-delayed fast dynamics being infinite-dimensional is that the stability of the relative equilibria with respect to fast perturbations (equivalently, the transverse stability of the critical manifold) is described by a quasipolynomial characteristic equation with infinitely many roots. 
Specifically, linearizing the fast subsystem around its relative equilibrium yields the characteristic equation:
\begin{align}
(\lambda+c_1)(\lambda+c_2)-c_1c_2e^{-2\lambda\tau}=0,
\label{eq:CharEq_adaptive}
\end{align}
where $c_1=\kappa_1\cos(\Omega_\ell\tau+\theta_\ell+\alpha)$ and $c_2=\kappa_2\cos(\Omega_\ell\tau-\theta_\ell+\alpha)$. The neutral eigenvalue $\lambda=0$ of \eqref{eq:CharEq_adaptive} always exists due to motion along the oscillation symmetry. 
For each selected relative equilibrium with $\Omega_\ell$ and $\theta_\ell$, Eq.~\eqref{eq:CharEq_adaptive} can be solved numerically to determine the equilibrium's stability properties. This allows the critical manifold $C_{crit}$ to be split into stable and unstable parts. 

\begin{figure*}
\centering		
\includegraphics[width=15.5cm]{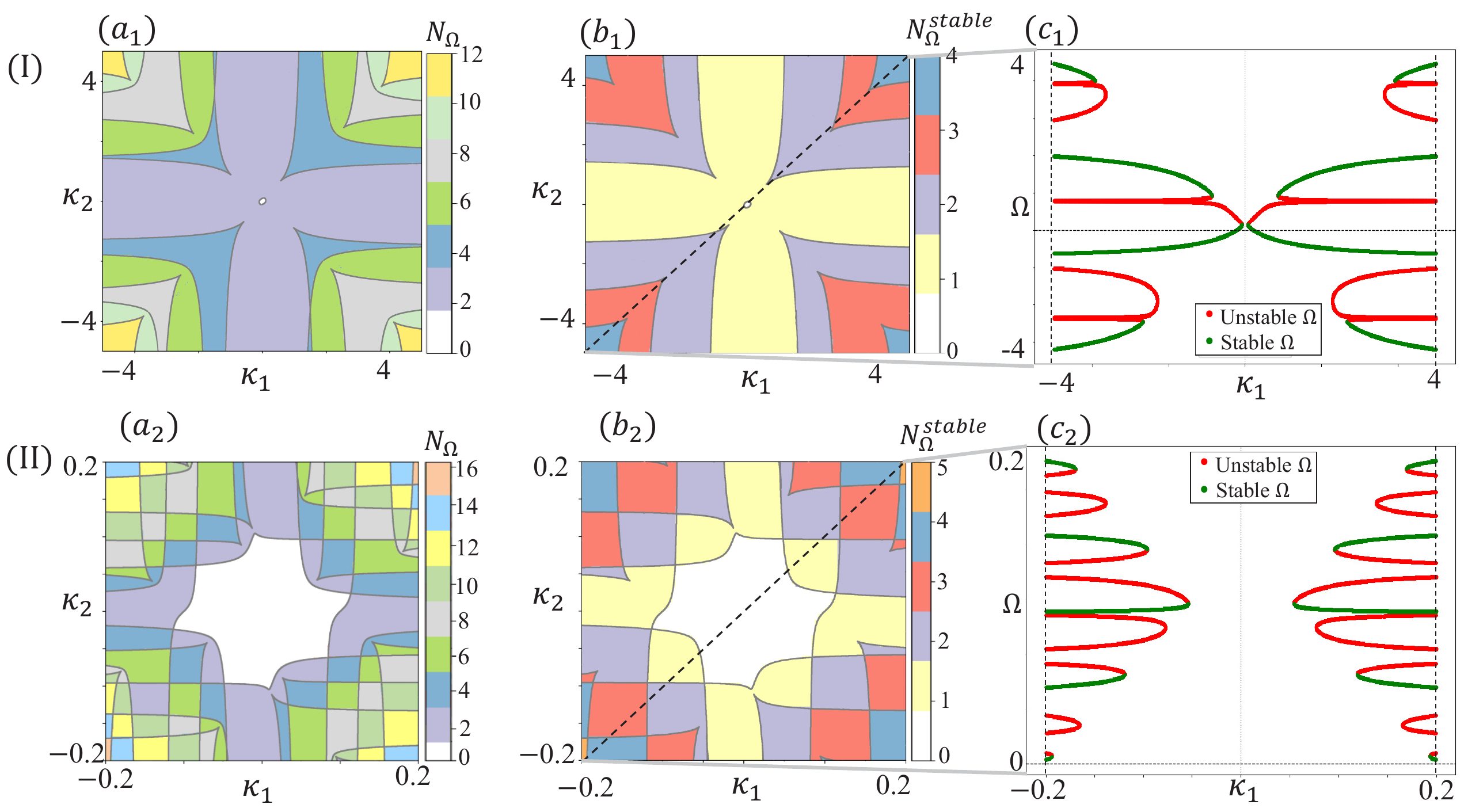}
	\caption{Critical-manifold structure for two delay values: (I) $\tau=1$ and (II) $\tau=40$. Panels $(a_1)$ and $(a_2)$ show the number of fast relative equilibria ($N_\Omega$) for each parameter pair $(\kappa_1,\kappa_2)$, while panels $(b_1)$ and $(b_2)$ show the number of stable fast relative equilibria  $N_\Omega^{\mathrm{stable}}$. 
    The color bars indicate distinct values of $N_\Omega$ and $N_\Omega^{\mathrm{stable}}$, and white regions contain no such oscillations. 
    Gray curves denote the projections of fold boundaries, where two fast relative equilibrium branches merge. 
    Panels $(c_1)$ and $(c_2)$ show cross-sections with $\kappa_1=\kappa_2$ of the critical manifolds, resolving overlapping projections into distinct branches $\Omega_\ell$ in the $(\kappa_1,\Omega)$-plane. The remaining parameters are $\omega_1=0.2$, $\omega_2=0.1$, and $\alpha={\pi}/{4}$,.
}
		\label{fig:CM_delay}
	\end{figure*}

To visualize the critical manifolds, Fig.~\ref{fig:CM_delay} shows projections of some of their properties onto the $(\kappa_1,\kappa_2)$ plane for $\tau=1$ (top row) and $\tau=40$ (bottom row). 
The first column of Fig.~\ref{fig:CM_delay} shows the number of sheets $N_\Omega$ of the critical manifold for different values of $(\kappa_1,\kappa_2)$, i.e., the number of relative equilibria. The second column of Fig.~\ref{fig:CM_delay} shows the number of stable sheets $N_{\Omega}^{\mathrm{stable}}$. 
To obtain this result, we solved Eq.~\eqref{eq:G_gamma_new} numerically for each fixed parameter pair $(\kappa_1,\kappa_2)$, and found the number $N_\Omega$ of distinct real solutions $\Omega_\ell$, $\ell=1,\cdots,N_\Omega$. 
For each relative equilibrium, we determine its spectral stability using Eq.~\eqref{eq:CharEq_adaptive}. 
After excluding the trivial eigenvalue, we classify the relative equilibrium as spectrally stable if all the remaining characteristic roots satisfy $\operatorname{Re}\lambda<0$. 
This defines the number of stable relative equilibria, $N_\Omega^{\mathrm{stable}}\le N_\Omega$. 

To illustrate the individual sheets contributing to this projection, we consider one-dimensional cross sections along the diagonal dashed lines given in  Figs.~\ref{fig:CM_delay}(b$_1$)--(b$_2$). 
The panels in Figs.~\ref{fig:CM_delay}(c$_1$)--(c$_2$) show 
branches in the $(\kappa_1,\Omega)$-plane, such that one can observe the frequencies $\Omega$ of the different coexisting sheets.
Each branch represents a one-dimensional section of a critical manifold sheet. 
Stable and unstable branches are distinguished by green and red colors. 
These branches show how the multiple roots for the same values of $(\kappa_1,\kappa_2)$ belong to different branches of the critical manifold. 
The branches meet at fold points, which correspond to lines in Figs.~\ref{fig:CM_delay}(a)--(b).

Relative equilibria of the full system \eqref{eq:phi1}--\eqref{eq:kappa2} are embedded into the critical manifolds $C_\ell$, see Appendix~\ref{subsec:Full_RE}. 

\section{Relative slow-fast periodic orbits: frequency bursts}
\label{sec:RPO_Phase_Locking}

In the previous section, we characterized the critical manifolds of the fast relative equilibria. We now relate these geometric structures to the FB relative periodic orbits introduced in Sec.~\ref{sec:relative_phase_locking}. Specifically, we show that the FB periodic motion decomposes into alternating slow and fast phases. The slow evolution is governed by the reduced dynamics close to stable critical manifold sheets, whereas the fast transitions connect different sheets. This slow-fast decomposition provides a geometric interpretation of the FB and explains the organization of the relative periodic orbits.

\subsection{Slow dynamics along the stable slow manifold sheets}
\label{sec:slow_dynamics_CMs}

We first derive the slow dynamics associated with each critical manifold sheet. 
Since the fast subsystem rapidly relaxes to a phase-locked relative equilibrium (one of the sheets $C_\ell$ shown in the $(\kappa_1,\kappa_2)$-plane in Fig.~\ref{fig:CM_delay}), the fast variables remain slaved to the slowly evolving adaptive variables. 
Consequently, the dynamics on each sheet are completely determined by the evolution of the adaptive variables, with the relative equilibrium parameters $(\kappa_1,\kappa_2)$ depending parametrically on the current location on the critical manifold. This yields a two-dimensional vector field on each sheet, which governs the slow drift between successive fast transitions.

Specifically, on the $\ell$-th sheet $ C_\ell$, the fast relative equilibrium is characterized by the oscillation frequency
$\Omega_\ell(\kappa_1,\kappa_2)$ and the relative phase difference $\theta_\ell(\kappa_1,\kappa_2)$. 
Substituting $\theta_\ell(\kappa_1,\kappa_2)$ into the adaptation equations \eqref{eq:kappa1}--\eqref{eq:kappa2} yields 
\begin{align}
\begin{split}
    \dot{\kappa}_1
&=
-\epsilon
\bigl(
\kappa_1-a_1\sin\theta_\ell(\kappa_1,\kappa_2)
\bigr),
\\
\dot{\kappa}_2
&=
-\epsilon
\bigl(
\kappa_2+a_2\cos\theta_\ell(\kappa_1,\kappa_2)
\bigr).
\label{eq:VectorField_sheet}
\end{split}
\end{align}
These equations define the slow evolution on $C_\ell$.
At each point $(\kappa_1,\kappa_2)$, the corresponding fast relative equilibrium determines $\theta_\ell(\kappa_1,\kappa_2)$ and the slow vector field \eqref{eq:VectorField_sheet} 
At each point of $C_\ell$, the vector $(\dot \kappa_1,\dot \kappa_2)$ according to Eq.~\eqref{eq:VectorField_sheet} gives the instantaneous velocity of the adaptive coupling variables rescaled by the factor $\epsilon$. Note that the vector field \eqref{eq:VectorField_sheet} is different on the different sheets $C_\ell$. 
The relative equilibria of the full slow-fast system, as derived in
\ref{subsec:Full_RE}, correspond to stationary points of Eq.~\eqref{eq:VectorField_sheet}.

\subsection{Slow-fast geometry of FB oscillations}
\label{subsec:RPOs_WJ}
In this subsection, we use the critical manifold  structure derived in
Sec.~\ref{subsec:CM_REs} and the slow vector fields \eqref{eq:VectorField_sheet} obtained in Sec.~\ref{sec:slow_dynamics_CMs} to interpret the FB oscillations introduced in Sec.~\ref{sec:relative_phase_locking}. 
We now focus on the geometric organization of these oscillations. 
Specifically, we identify slow trajectories that remain close to stable critical manifold sheets and examine the fast relative phase transitions.

\begin{figure*}
		\centering	
\includegraphics[width=0.4\textwidth]{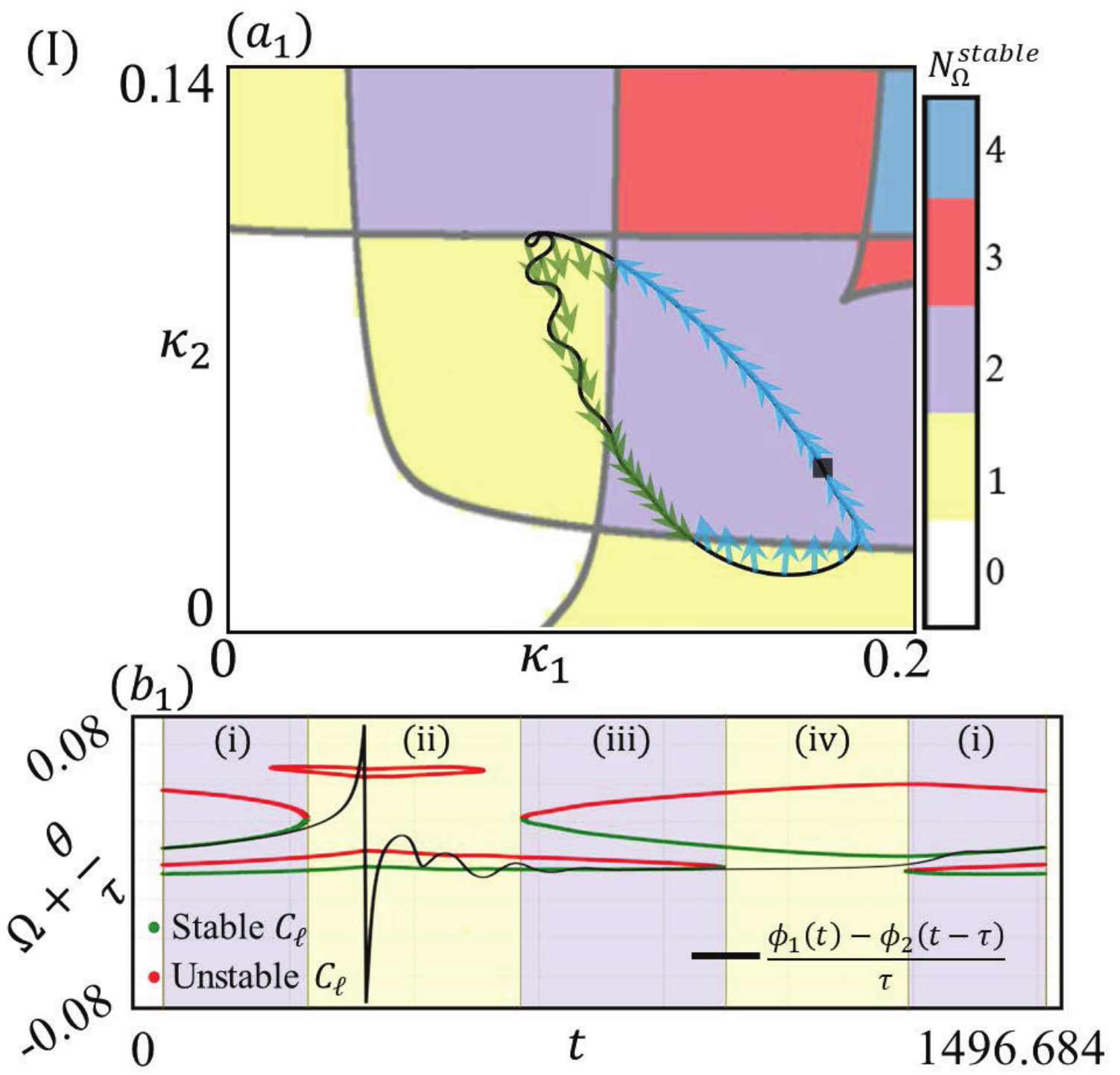}
\includegraphics[width=0.4\textwidth]{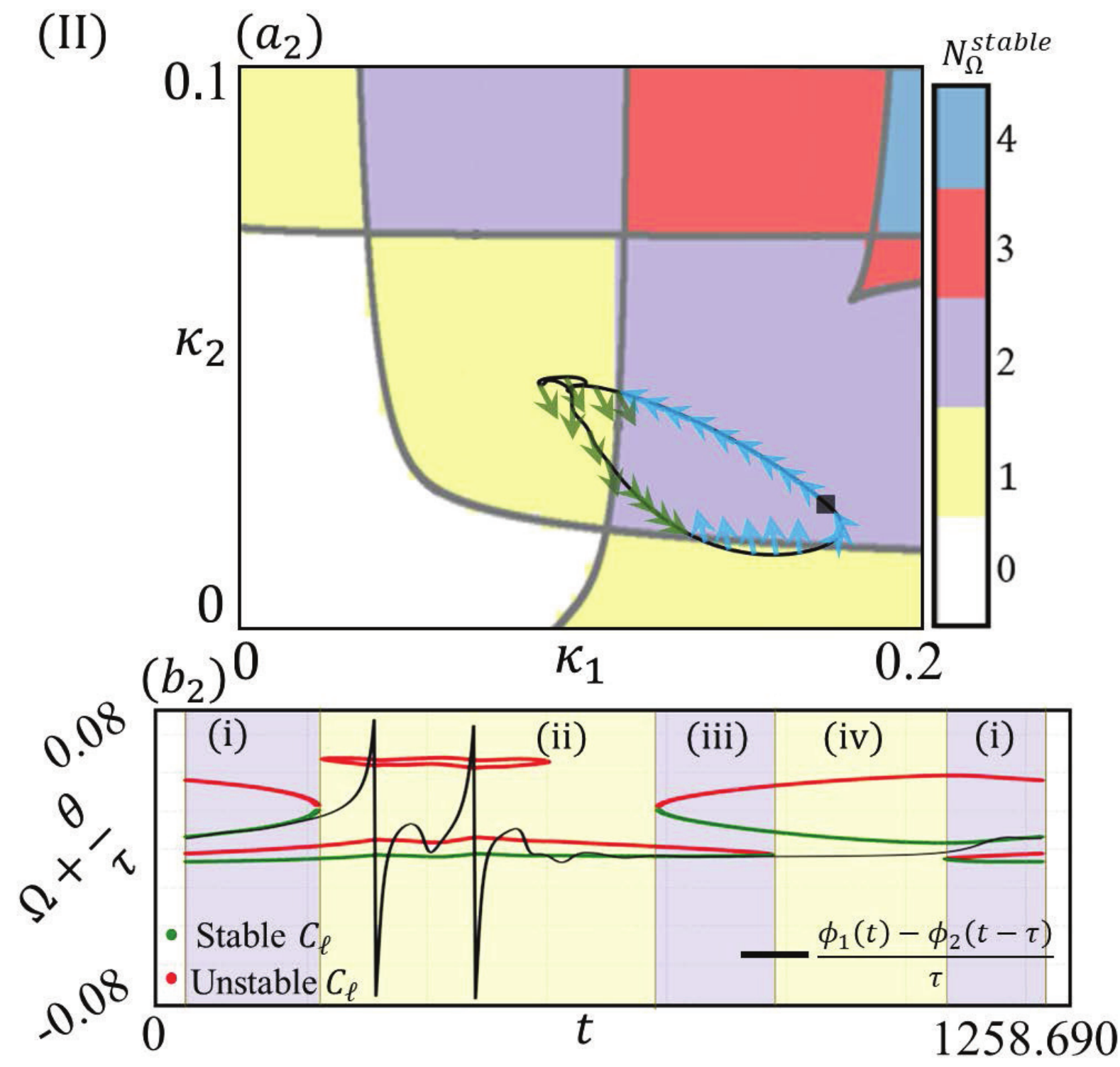}
\includegraphics[width=0.4\textwidth]{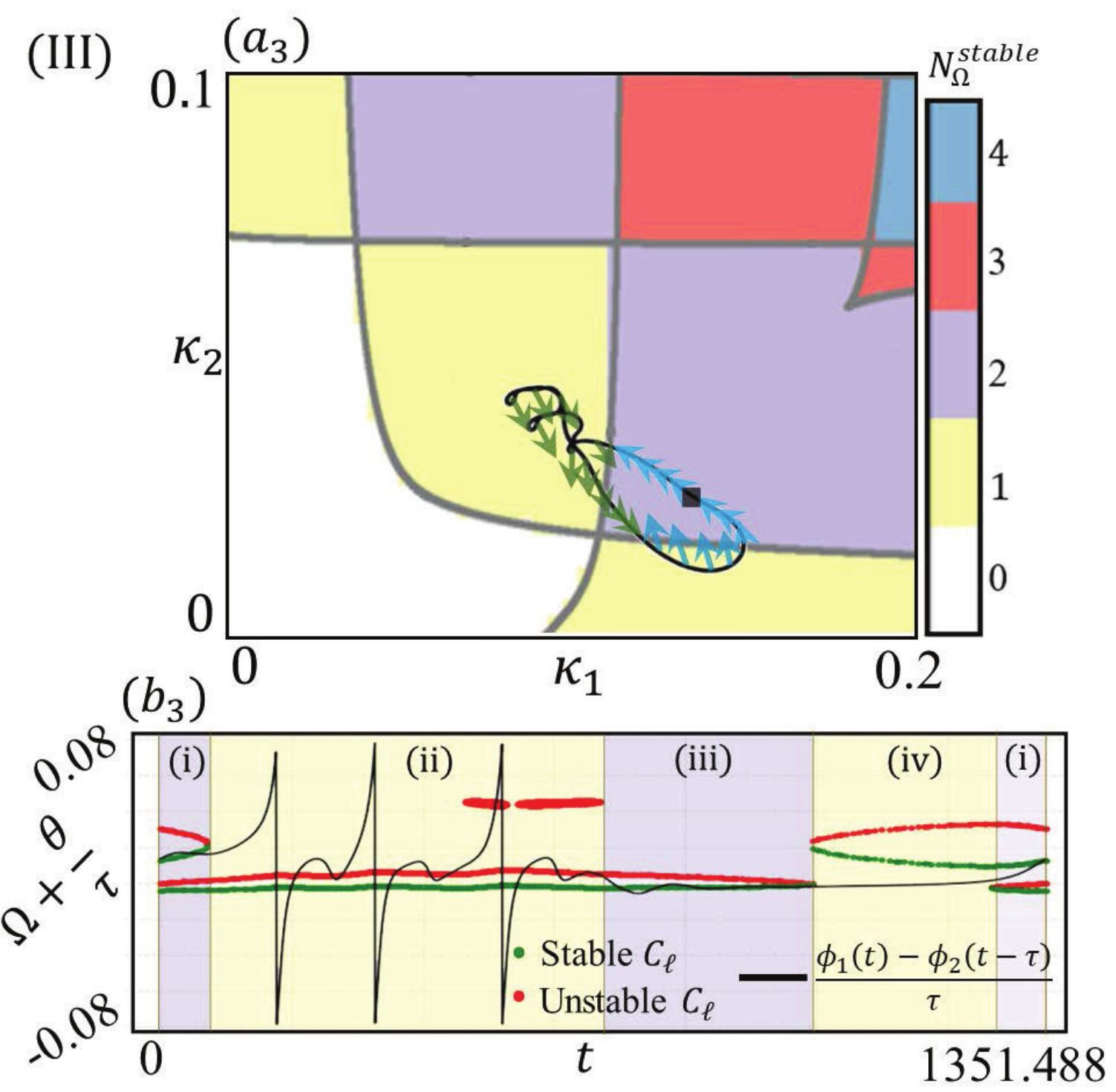}
\caption{Representative FB orbits with different number of frequency spikes $m$.
Cases (I)--(III) correspond to $m=1$, $2$, and $3$, with $(a_1,a_2)=(0.5,0.4)$, $(0.3,0.25)$, and $(0.25,0.26)$, respectively.
Panels $(a)$ show the orbits projected onto the
$(\kappa_1,\kappa_2)$-plane and superimposed on the critical-manifold
structure shown in Fig.~\ref{fig:CM_delay}(II)(b).
The black closed curve represents one period of the slow evolution,
and the black square marks its initial and final point.
Blue and green arrows indicate the two stable reduced vector fields.
Panels $(b_1)$--$(b_3)$ show the corresponding evolution of the delayed relative phase 
$\bigl[\phi_1(t)-\phi_2(t-\tau)\bigr]/\tau$.
Green and red lines show stable and unstable sheets of the critical manifold of relative equilibria, respectively, i.e., the values of  $\Omega_\ell(\kappa_1(t),\kappa_2(t))+\theta_\ell(\kappa_1(t),\kappa_2(t))/\tau$ are plotted that correspond to the value of $\bigl[\phi_1(t)-\phi_2(t-\tau)\bigr]/\tau$ when evaluated at the critical manifolds. 
The colored vertical bands indicate the time intervals spent in the
corresponding critical-manifold domains and match the background regions
in panels $(a)$.
The remaining parameters are
$\alpha={\pi}/{4}$,
$\omega_1=0.2$, $\omega_2=0.1$,
$\epsilon=4\times10^{-4}$, and $\tau=40$.
} \label{fig:RPO_onetothree_jumps} 
\end{figure*}

Figure~\ref{fig:RPO_onetothree_jumps} shows three representative FB oscillations with  $m=1$, $2$, and $3$, respectively. 
Panels $(a_1)$--$(a_3)$ display their projections into the $(\kappa_1,\kappa_2)$-plane of slow variables. 
The black curve shows the periodic orbit.
The gray curves mark the folds of the critical manifolds, at which the fast jumps are expected. 
The arrows along the black curves indicate the directions of the slow vector fields \eqref{eq:VectorField_sheet} along the sheet of the critical manifold, which is closest to the periodic trajectory. 
The different colors of the vector field correspond to the two distinct stable critical manifold sheets followed by the trajectory (here blue and green).

Panels $(b_1)$--$(b_3)$ in Fig.~\ref{fig:RPO_onetothree_jumps} show the values of 
$(\phi_1(t)-\phi_2(t-\tau))/\tau$ of the FB oscillations (black curve), as an appropriate projection of the fast dynamics. In this projection a relative equilibrium of the full system \eqref{eq:phi1}--\eqref{eq:kappa2} would be a constant (a horizontal line, not shown). The FB oscillation (in black) follows the 
projection of the critical manifolds (green for stable, red for unstable parts) for a long time, interrupted by frequency jumps: the FB oscillation in panel $(b_1)$ has $m=1$ jump, in panel $(b_2)$ it has $m=2$ jumps, in panel $(b_3)$ it has $m=3$ jumps.
Specifically, the value of $(\phi_1(t)-\phi_2(t-\tau))/\tau$, evaluated at the critical manifolds, gives 
\begin{equation}
\left.\frac{\phi_1(t)-\phi_2(t-\tau)}{\tau} \right|_{C_\ell}
=
\Omega_\ell\bigl(\kappa_1(t),\kappa_2(t)\bigr)
+
\frac{
\theta_\ell\bigl(\kappa_1(t),\kappa_2(t)\bigr)
}{\tau},
\label{eq:cross_section_sheet}
\end{equation}
hence, this quantity is compared with the orbit. 

Panels $(a_1)$--$(a_3)$ and $(b_1)$--$(b_3)$ in Fig.~\ref{fig:RPO_onetothree_jumps} describe complementary aspects of the slow-fast dynamics with frequency jumps. 
While panels $(a)_1$--$(a_3)$ show the slow evolution of the adaptive variables, panels $(b_1)$--$(b_3)$ provide the fast bursts as well as the slow manifold branch $C_\ell$ near which the slow evolution takes place.
The stable critical manifold sheets contain the slow segments of the orbit, while fast
transitions between these sheets generate the frequency jumps.

For each of the $3$ cases in Fig.~\ref{fig:RPO_onetothree_jumps} the orbit's evolution can be split into four qualitatively different periodically repeating phases:

(i) Slow motion along a sheet $C_\ell$ of the slow manifold of relative equilibria, with the vector field on the manifold tangential to the orbit (blue arrows in $(a_1)$--$(a_3)$). In panels $(b_1)$--$(b_3)$, the orbit is aligned with the green line of the stable critical manifold in this phase. During this evolution, the fast frequency of the oscillation is slowly drifting according to $\Omega_\ell(\kappa_1(t),\kappa_2(t))$.  

(ii) Due to the fold of the critical manifold, the orbit undergoes a jump, leading to fast oscillations of $\phi_1-\phi_2$ and fast spikes in the instantaneous frequency. The orbit with $m=1$ spike (in panel $(b_1)$), $m=2$ spikes (in panel $(b_2)$), or $m=3$ spikes (in panel $(b_3)$) in the burst makes $m$ such fast oscillations before converging to another sheet $C_\ell$ of the stable part of the slow manifold of relative equilibria.
This phase (ii) is critical for the emergence of frequency bursts. It arises from the interplay between the time-delay $\tau$ and the timescale separation $\varepsilon $. Indeed, for a large $\tau$, the transverse modes of the stable critical manifold possess attraction rates that scale as $1/\tau$ \cite{Lichtner2011}, which restricts the attraction rate to the critical manifold and enables such oscillations.
The corresponding relaxation time therefore scales as $\tau$, during which the adaptive variables change by an amount of order $\epsilon\tau$.
This is also the reason why such oscillations disappear as $\epsilon$ decreases.
For fixed $\tau$, the adaptive variables become effectively frozen during the fast excursion as $\epsilon\tau$ decreases, and the orbit is recaptured by another stable sheet before such oscillations can develop
(see the following sections).

(iii) Slow motion along another manifold of relative equilibria. 

(iv) A longer monotone transition occurs to another stable manifold of relative equilibria. As in case~(ii), after the critical manifold disappears in a fold, the repulsion from the `ghost' attractor is relatively slow, resulting in an extended transition period. Unlike case~(ii), this transition period does not produce frequency spikes.

In summary, Fig.~\ref{fig:RPO_onetothree_jumps} 
illustrates the connection between the static critical manifold structure and the FB dynamics. 
The stable sheets, together with their associated vector fields, organize the slow segments. Departures from these sheets at fold bifurcation boundaries, give rise to fast frequency jumps. 
In the examples shown in Fig.~\ref{fig:RPO_onetothree_jumps}, the
orbits involve only two phase-locked sheets, corresponding to (i) and
(iii), although more than two sheets participate in
an FB itinerary.
\section{Parameter dependence and multistability of FB oscillations} \label{sec:parameter_organization_RPOs}

We now study how FBs depend on system parameters.
First, we vary the adaptive-coupling strengths $(a_1,a_2)$ to identify regions, where stable FB oscillations with different burst sizes $m$ exist.
We then use the DDE-BifTool \cite{sieber2014ddebiftool} to continue representative FB oscillations in the $(\epsilon, \tau)$-plane and determine their stability regions and bifurcation boundaries.
Finally, we demonstrate FB multistability, when stable FBs with different $m$ coexist.

\subsection{Parameter regions for FB} \label{subsec:winding_domains_continuation}
 
\begin{figure}
\includegraphics[width=\linewidth]{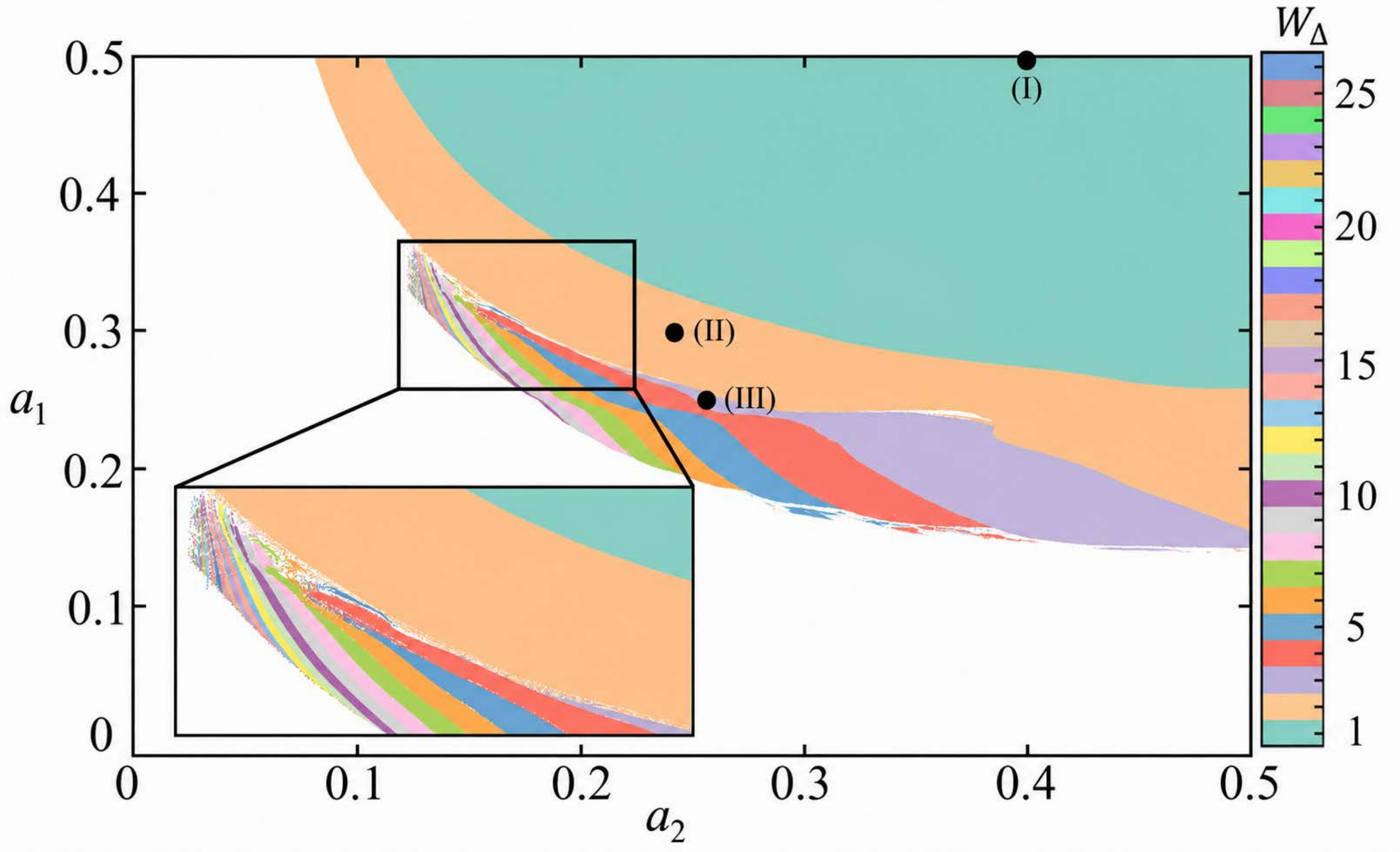} 
\caption{FB oscillations in the $(a_1,a_2)$ parameter plane. The oscillations are obtained starting from the fixed initial condition \eqref{eq:history_H0}.
Colors indicate the integer value $W_{\Delta}(T_m)=m$ defined in \eqref{eq:WindingJump}, while white regions denote parameter values for which no FB  oscillation was detected. The fixed parameters are $\alpha={\pi}/{4}$, $\beta_1=0$, $\beta_2=-{\pi}/{2}$, $\omega_1=0.2$, $\omega_2=0.1$, $\tau=40$, and $\epsilon=4\times10^{-4}$. Points (I)--(III) correspond to the representative solutions in Fig.~\ref{fig:RPO_onetothree_jumps}. The map therefore represents the attracting state selected by this fixed numerical scan and does not exclude coexisting attractors associated with other initial histories. The inset enlarges a region where FB domains with different $m$ appear to accumulate.} \label{fig:WindingDiff_ab_map}
 \end{figure}

To demonstrate the robustness and abundance of FB, we scan the parameter plane $(a, b)$ of adaptive coupling strengths and compute the asymptotic value of $m = W_{\Delta}(T_m)$ for each parameter pair with the fixed initial condition 
\begin{equation}
\label{eq:history_H0}
\begin{aligned}
& \phi_1(t)=0,~
\phi_2(t)=0,~
t\in[-\tau,0],\\
& \kappa_1(t)=0.2,~
\kappa_2(t)=0.1.
\end{aligned}
\end{equation} 
We observe that the $(a_1,a_2)$ plane is divided into several regions (see Fig.~\ref{fig:WindingDiff_ab_map}), each with a distinct integer value of $m$. 
Therefore, the burst size can be controlled by adjusting the parameters of the adaptations. Interestingly, the inset shows that a small region of parameters exhibits multiple nearby regions with large $m$, indicating a sensitive dependence on parameter changes. Regions that appear `noisy' correspond to multistability, meaning that even small parameter variations can lead to different attractors. This multistability will be discussed in more detail later. 

\begin{figure}
\centering
\includegraphics[width=0.56\linewidth]{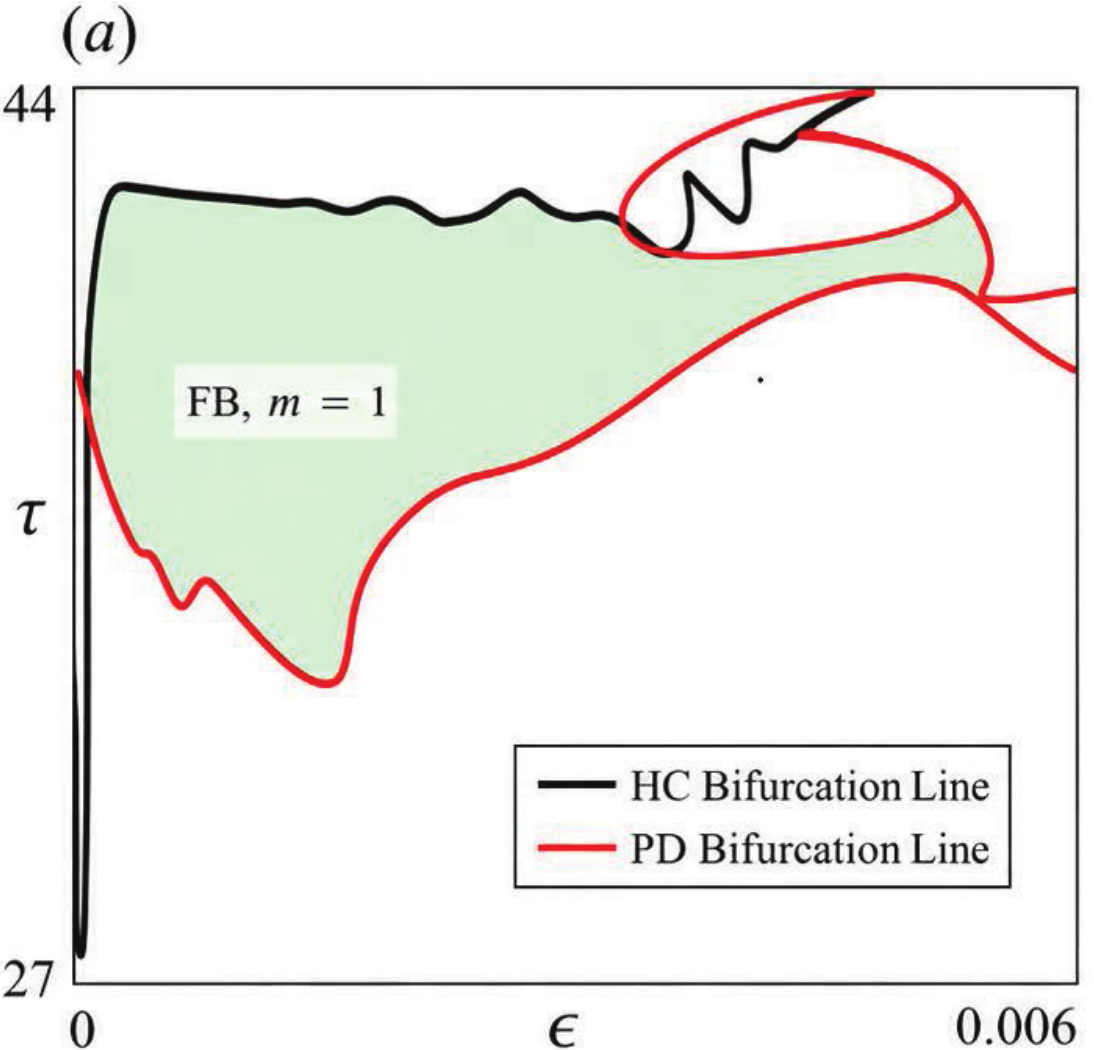}\\
\includegraphics[width=0.48\linewidth]{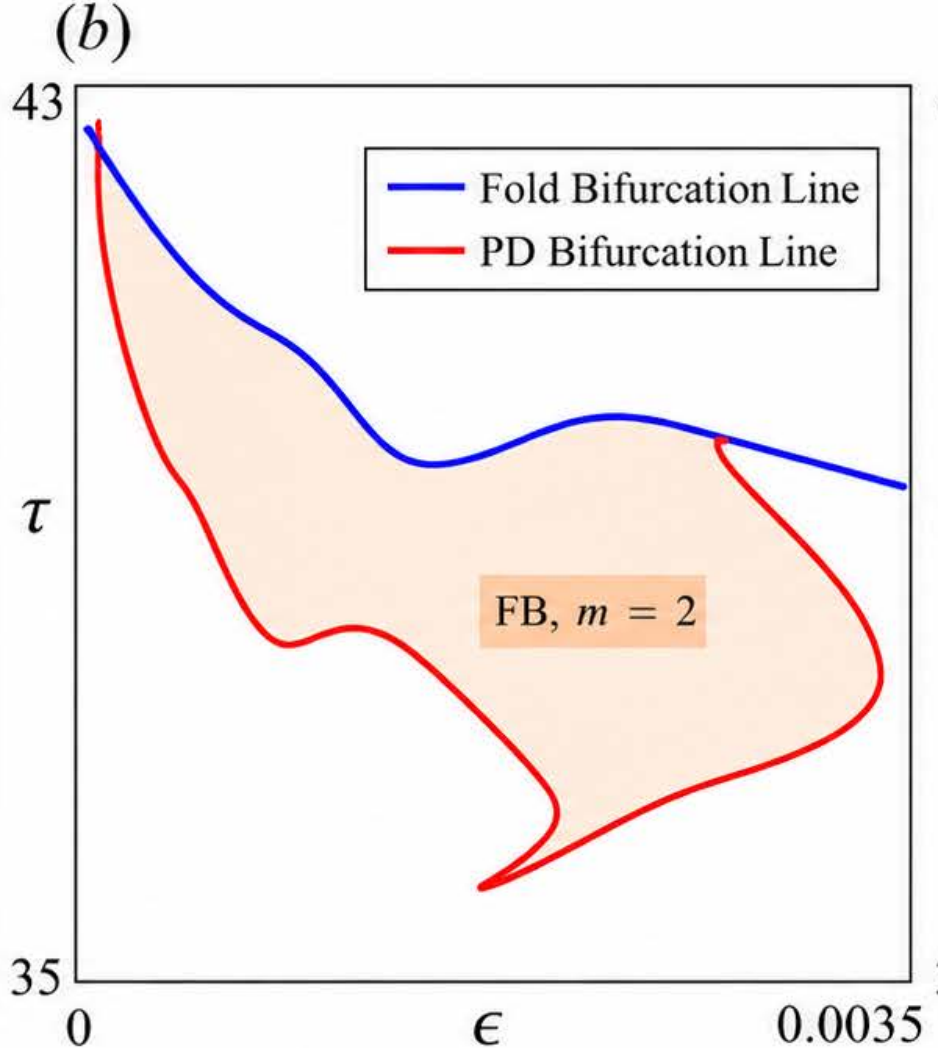}
\includegraphics[width=0.5\linewidth]{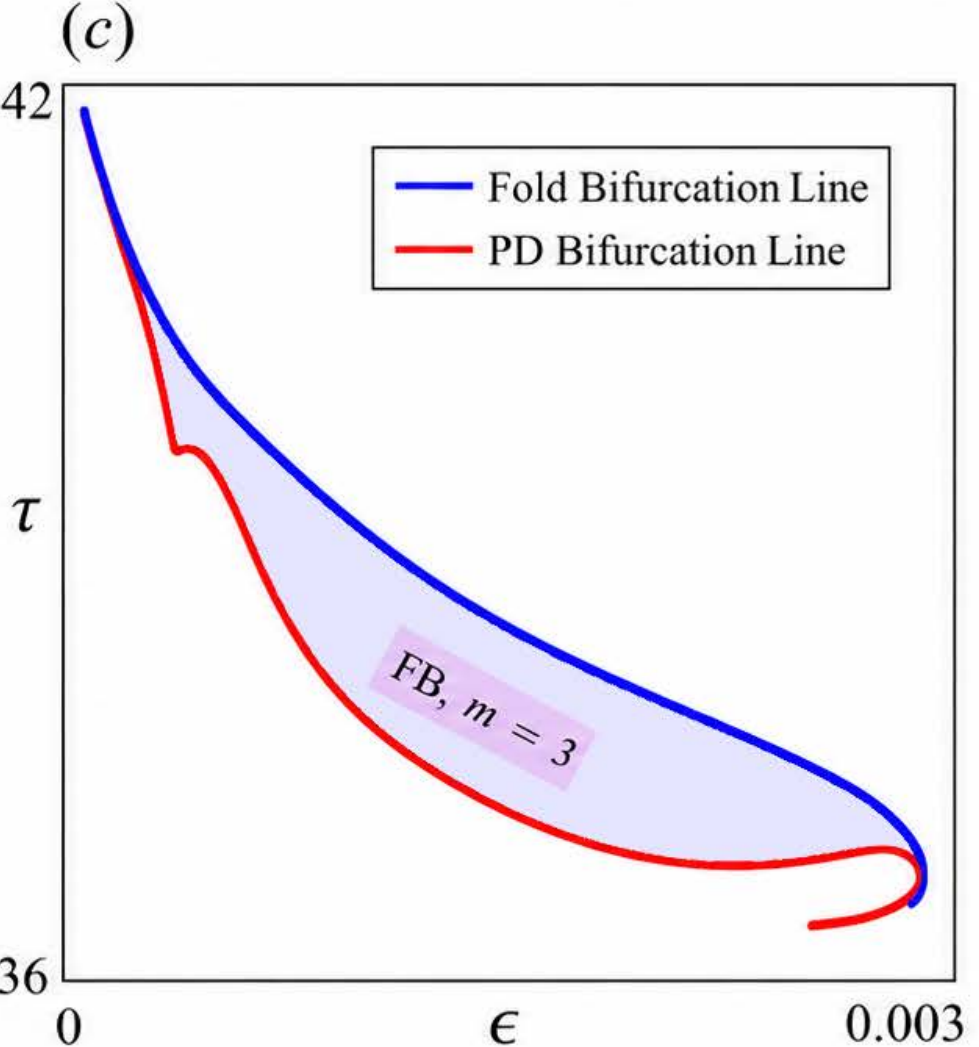}
    \caption{
    Numerical continuation of three representative FB orbits using DDE-BifTool.
    Panels (a)--(c) show the corresponding stability diagrams in the
    $(\epsilon,\tau)$ plane for burst size $m=1$, $m=2$, and $m=3$,
    respectively.
    The shaded regions indicate parameter domains in which the corresponding
    FB orbits are stable.
    Red, blue, and black curves denote period-doubling (PD), homoclinic (HC),
    and fold bifurcations, respectively.
    }
    \label{fig:RPO_continuation}
\end{figure}

Next, we perform a numerical bifurcation analysis of the FB orbits with $m=1,2,3$ with respect to the adaptation rate $\epsilon$ and the time delay $\tau$, as shown in Fig.~\ref{fig:RPO_continuation} using the DDE-BifTool \cite{sieber2014ddebiftool}.
Panels $(a)$--$(c)$ show the corresponding two-parameter continuation diagrams in the $(\epsilon,\tau)$-plane for $m=1$, $m=2$, and $m=3$, respectively.
The shaded regions indicate the parameter domains in which the corresponding FB orbit is stable.
The red curves denote period-doubling bifurcation boundaries, whereas the blue ones correspond to fold bifurcations.
In panel $(a)$, the black curve denotes a homoclinic bifurcation boundary. 

\subsection{Multistability of FB}
\label{subsec:history_dependent_multistability}
Here we demonstrate the multistability of FB. 
This is associated with the presence of multiple stable sheets $C_\ell$ of the critical manifold, which can organize distinct coexisting slow-fast itineraries.
Since the initial value problem for delayed systems requirse an initial history on the interval
$t\in[-\tau,0]$, different history functions may belong to different basins of attraction, even when all system parameters are fixed. 
Consequently, the same adaptive-coupling parameters may support the multistability, i.e. the coexisting of stable FB periodic orbits with different number of spikes in the burst.

\begin{figure}
    \centering    \includegraphics[width=0.92\linewidth]{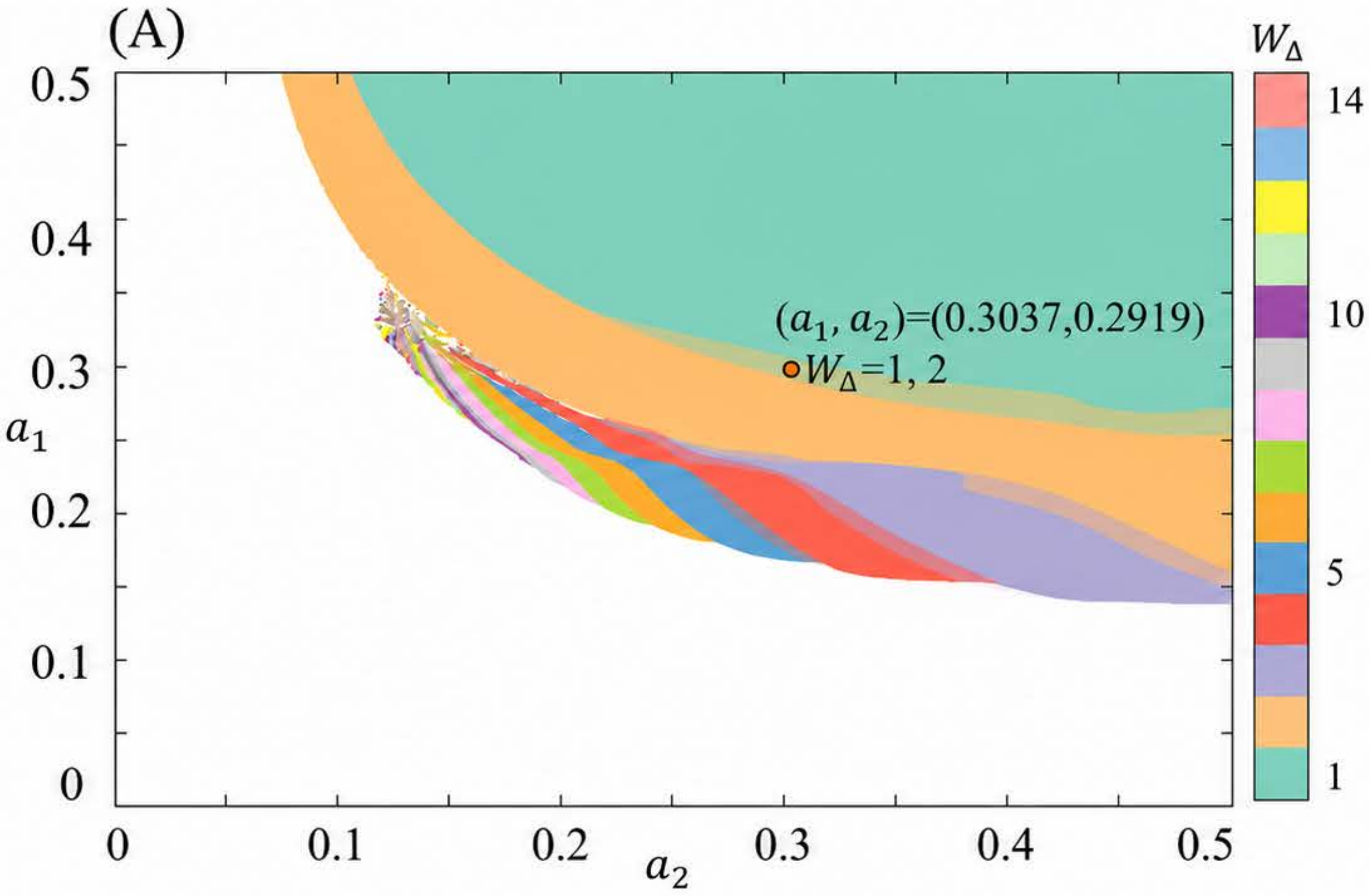}
\includegraphics[width=0.92\linewidth]{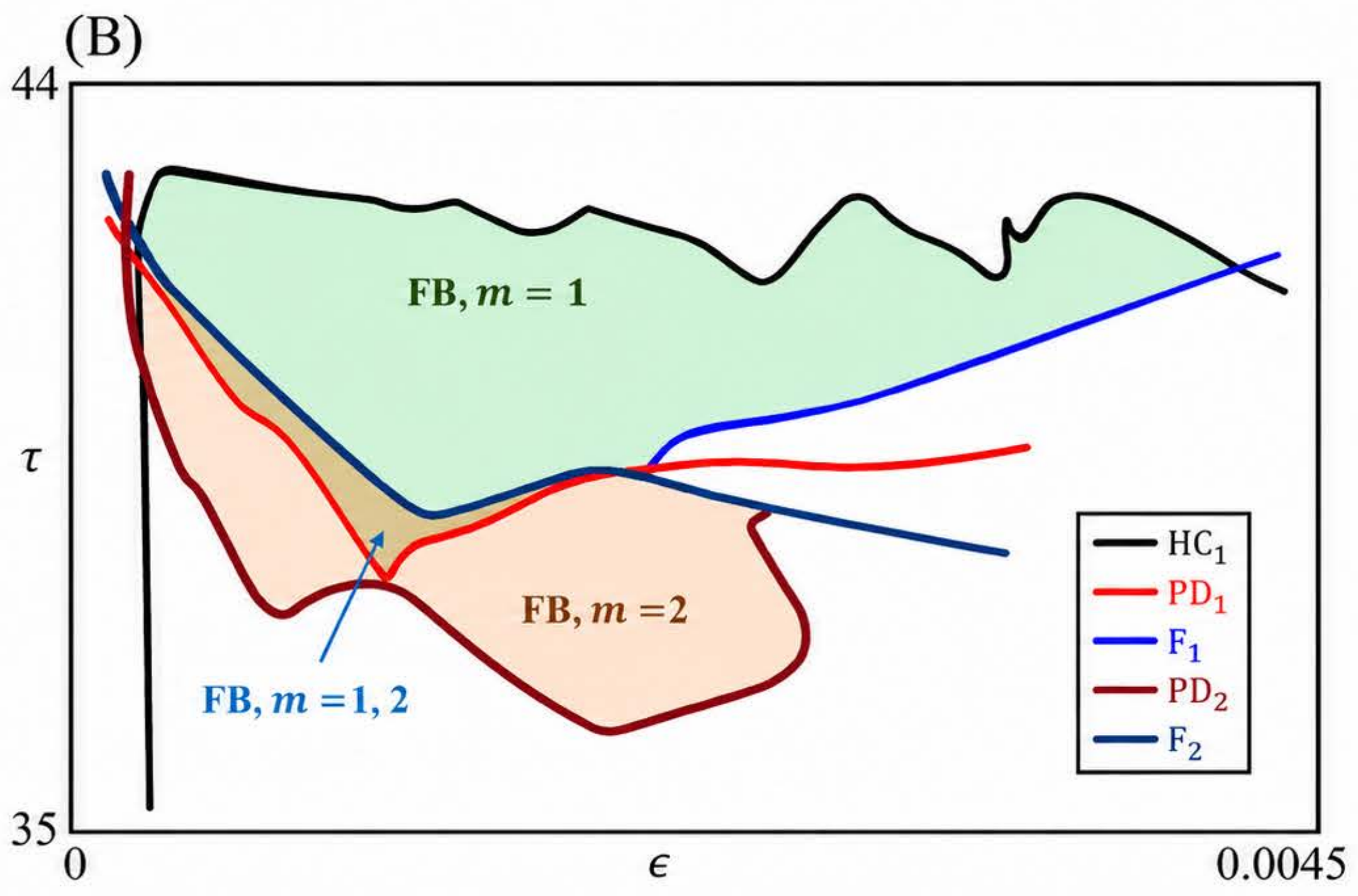}
    \caption{Initial-history-dependent multistability and its continuation in parameter space.
(A) Superposition of the distributions of the relative winding $m=W_{\Delta}(T_m)$ in the $(a_1,a_2)$-plane obtained from the two initial histories, as defined in Eqs.~\eqref{eq:history_H0} and \eqref{eq:history_H6}.
Different colors indicate FB orbits with different integer values of $m$.
At the marked point $(a_1,a_2)=(0.3037,0.2919)$, the histories  converge to FB orbits with $W_\Delta=1$ and $W_\Delta=2$, respectively.
(B) Continuation of these two FB oscillation families in the $(\epsilon,\tau)$-plane at the marked values of $(a_1,a_2)$.
The green and orange regions correspond to stable FB orbits with $m=1$ and $m=2$, respectively, while their overlap corresponds to the coexistence region labeled $\mathrm{FB},\,m=1,2$.
The curves $\mathrm{PD}_m$ and $\mathrm{F}_m$ denote period-doubling and fold bifurcations of the $m$ family, respectively, whereas $\mathrm{HC}_1$ denotes the homoclinic bifurcation associated with the $m=1$ family.
Other parameters are $\alpha={\pi}/{4}$, $\beta_1=0$, $\beta_2=-{\pi}/{2}$, $\omega_1=0.2$, and $\omega_2=0.1$.
    }
\label{fig:history_multistability_combined}
\end{figure}

To illustrate the multistability, we use two initial history functions. The first one is a non-winding constant history \eqref{eq:history_H0}.
The second history has the following form
\begin{align}
\begin{cases}
\phi_1(t)
=
\frac{\pi}{4}+\frac{3}{20} t
+\dfrac{1}{2}\left(\pi+12\pi \dfrac{t}{\tau}\right),\\[2mm]
\phi_2(t)
=
\frac{\pi}{4}+\frac{3}{20}t
-\dfrac{1}{2}\left(\pi+12\pi \dfrac{t}{\tau}\right),\\
\kappa_1(t)=0.08,\quad
\kappa_2(t)=0.16,
\end{cases}
\qquad t\in[-\tau,0],
\label{eq:history_H6}
\end{align}
which has nonzero relative winding over the delay interval $[(\phi_1(0)-\phi_2(0))-(\phi_1(-\tau)-\phi_2(-\tau)]/(2\pi)=6$.
Figure~\ref{fig:history_multistability_combined}(A) shows the superposition of the stability diagrams, obtained from the two prescribed histories, Eqs.~\eqref{eq:history_H0} and \eqref{eq:history_H6}, in the $(a_1,a_2)$-plane.
The colors indicate the resulting integer relative winding $m=W_{\Delta}(T_m)$.
When both histories yield different values of $m$ at the same parameter pair, distinct FB orbits can be selected by different initial histories, indicating multistability.
The marked point $(a_1,a_2)=(0.3037,0.2919)$ provides a representative example.
For this same parameter pair, the oscillation starting from the history Eq.~\eqref{eq:history_H0} 
converges to a FB orbit with $W_{\Delta}=1$, whereas
the oscillation starting from the history defined in Eq.~\eqref{eq:history_H6} converges to a different FB orbit with $W_{\Delta}=2$.

Figure~\ref{fig:history_multistability_combined}(B) shows that the stable regions of  FB orbits with $m=1$ and $m=2$ overlap over a finite domain in the $(\epsilon,\tau)$-plane.
To obtain the diagram in Fig.~\ref{fig:history_multistability_combined}(B), we continue both stable FB orbits related to the marked point $(a_1,a_2)=(0.3037,0.2919)$, while $\epsilon$ and $\tau$ are varied.
$\mathrm{PD}_m$ and $\mathrm{F}_m$ denote the period-doubling and fold bifurcation curves of the FB orbits family with $m$ spikes, respectively, while $\mathrm{HC}_1$ denotes a homoclinic bifurcation associated with the $m=1$ family.

\section{Conclusion}\label{sec:conclusions}
In this work, we have studied the frequency bursting (FB) and near synchrony quantized detuning in adaptively coupled phase oscillators with time delay. 
These dynamics are organized by the critical manifolds of the fast relative equilibria.
We further explored the geometric properties, stability, stable continuation, and multistability of FB oscillations.
Different FB oscillations are distinguished by the number of spikes in the frequency burst. 
Despite the infinite-dimensional phase space induced by the time delay,
the FB dynamics has an effective low-dimensional organization. 
For two oscillators, the slow evolution follows the two-dimensional critical-manifold sheets for long times between fast transitions.

We find the following main properties of FB oscillations: 
\begin{itemize}
    \item The FB oscillations have the form of periodic bursts of the relative frequencies $\dot \phi_1 - \dot\phi_2$ with $m$ spikes in the burst.
    \item The difference of their mean frequencies $\Omega_1-\Omega_2$ is proportional to $\epsilon$, the small parameter characterizing the inverse of the adaptation timescale. 
    \item The mean frequency difference $\Omega_1-\Omega_2$ is quantized: $\Omega_1-\Omega_2 \sim m \epsilon$, $m=1,2,3,\cdots$. We call it \textit{near synchrony quantized detuning}.
\end{itemize}

For systems with frequency bursts, the ratio of the mean frequencies alone cannot provide a reliable characterization of the observed frequency-locked states. 
The ratio $\Omega_1/\Omega_2=1/(1+2\pi m/(\Omega T_m))$ for the FB oscillations need not be rational, and may in low-accuracy environments be indistinguishable from $1$.
We characterized the dynamics in terms of the relative winding number. 
Over one adaptive period, this accumulated relative winding is an integer, $W_\Delta(T_m)=m$. 
It quantifies the phase-shift per period between the oscillators and distinguishes different FB oscillations.
For the FB oscillations considered here, where phase jumps occur only during one transition. 
At the same time, $m$ coincides with the burst size.

The parameter organization analysis and the continuation of representative relative periodic orbits further show that the  frequency burst states are robust and that there can be many coexisting frequency burst states as the slow manifold has many overlapping stable sheets. 

In summary, these propose a general framework for analyzing delayed adaptive systems, which are characterized by slow `finite-dimensional' adaptation and fast `delay-driven' high-dimensional dynamics. 
An important open question concerns the mechanisms underlying the loss of stability of periodic frequency-bursting states and their transition to chaotic bursting.

\begin{acknowledgments}
This work was supported by  Taighde {\'E}ireann--Research Ireland (Grant No.~FFPA/12066).
 We also acknowledge valuable discussions with Matthias Wolfrum and Oleksandr Burylko.
\end{acknowledgments}

\appendix

\section{Rescaled system}\label{sec:nondim}
The frequency difference $\Delta\omega=\omega_2-\omega_1>0$ defines a natural frequency scale for the system. Here, the phase variables describe rotators that evolve on the circle and can perform complete rotations. To make the natural frequency scale explicit, we first introduce a rotating frame,
$\widetilde{\phi}_j=\phi_j-\omega_1 t$, which transforms the natural frequencies from
$(\omega_1,\omega_2)$ to $(0,\Delta\omega)$. Rescaling time subsequently as $\widetilde{t}=\Delta\omega t$ amounts to measuring frequencies in units of $\Delta\omega$, so that the natural frequencies are normalized to $\widetilde{\omega}_1=0$ and $\widetilde{\omega}_2=1$.
Under these transformations, the remaining parameters are rescaled according to $\widetilde{\tau}=\Delta\omega\tau$,
$\widetilde{\epsilon}=\epsilon/\Delta\omega$, $\widetilde{a}_i=a_i/\Delta\omega$,
and $\widetilde{\kappa}_i=\kappa_i/\Delta\omega$.
Moreover, the rotating-frame transformation introduces an additional phase shift $\omega_1\tau$ in the delayed coupling, leading to
$\widetilde{\alpha}=\alpha+\omega_1\tau$, whereas the adaptation phase
shifts $\beta_i$ remain unchanged because the adaptation laws depend only
on instantaneous phase differences. Hence, for $\Delta\omega>0$, the
system can be equivalently represented in this normalized form without loss of generality.
The subsequent analysis is carried out in the original parametrization, while the normalized form above serves to identify the natural scaling of the system.

In terms of the rescaled variables, the system can therefore be written as the follow form:
\begin{align}
\frac{d\widetilde{\phi}_1}{d\widetilde{t}}
&=
-\widetilde{\kappa}_1
\sin\left[
\widetilde{\phi}_1(\widetilde{t})
-\widetilde{\phi}_2(\widetilde{t}-\widetilde{\tau})
+\widetilde{\alpha}
\right], \\
\frac{d\widetilde{\phi}_2}{d\widetilde{t}}
&=
1-\widetilde{\kappa}_2
\sin\left[
\widetilde{\phi}_2(\widetilde{t})
-\widetilde{\phi}_1(\widetilde{t}-\widetilde{\tau})
+\widetilde{\alpha}
\right], \\
\frac{d\widetilde{\kappa}_1}{d\widetilde{t}}
&=
-\widetilde{\epsilon}
\left[
\widetilde{\kappa}_1
-\widetilde{a}_1
\sin\left(
\widetilde{\phi}_1-\widetilde{\phi}_2+\beta_1
\right)
\right], \\
\frac{d\widetilde{\kappa}_2}{d\widetilde{t}}
&=
-\widetilde{\epsilon}
\left[
\widetilde{\kappa}_2
-\widetilde{a}_2
\sin\left(
\widetilde{\phi}_2-\widetilde{\phi}_1+\beta_2
\right)
\right].
\end{align}
Thus, for $\Delta\omega>0$, the system can be equivalently represented with normalized natural frequencies $(0,1)$ without loss of generality. The subsequent analysis is carried out in the original parametrization, while the normalized form above serves to identify the natural scaling of the system.

\section{Relative equilibria of the full system}
\label{subsec:Full_RE}

We now derive the relative equilibria of the full slow-fast system \eqref{eq:phi1}--\eqref{eq:kappa2}. 
These solutions correspond to phase-locked states, where both phases oscillate with the common frequency $\phi_1=\Omega t$ and $\phi_2=\Omega t -\theta$, while the adaptive coupling variables $\kappa_1$ and $\kappa_2$ remain stationary. 
We consider the Hebbian-like adaptation $\beta_2=-\pi/2$. Then, the stationarity of $\kappa_i$, i.e., $\dot \kappa_i=0$, leads to 
\begin{align*}
\kappa_1=a_1\sin\theta,\quad\kappa_2=-a_2\cos\theta.
\end{align*}
Substituting these values of $\kappa_1$ and $\kappa_2$ into Eqs.~\eqref{eq:Fast_phi1}--\eqref{eq:Fast_phi2}, we obtain the equations for $\Omega$ and $\theta$:
\begin{subequations}
\label{eq:Fast_Omega}
\begin{align}
\Omega&=\omega_{1}-a_1\sin\theta\sin\left(\theta+\Omega\tau+\alpha\right), 
\label{eq:Omega1}\\
\Omega&=\omega_{2}+a_2\cos\theta\sin\left(-\theta+\Omega\tau+\alpha\right). 
\label{eq:Omega2}
\end{align}
\end{subequations}

\begin{figure}
\centering
\includegraphics[width=0.92\linewidth]{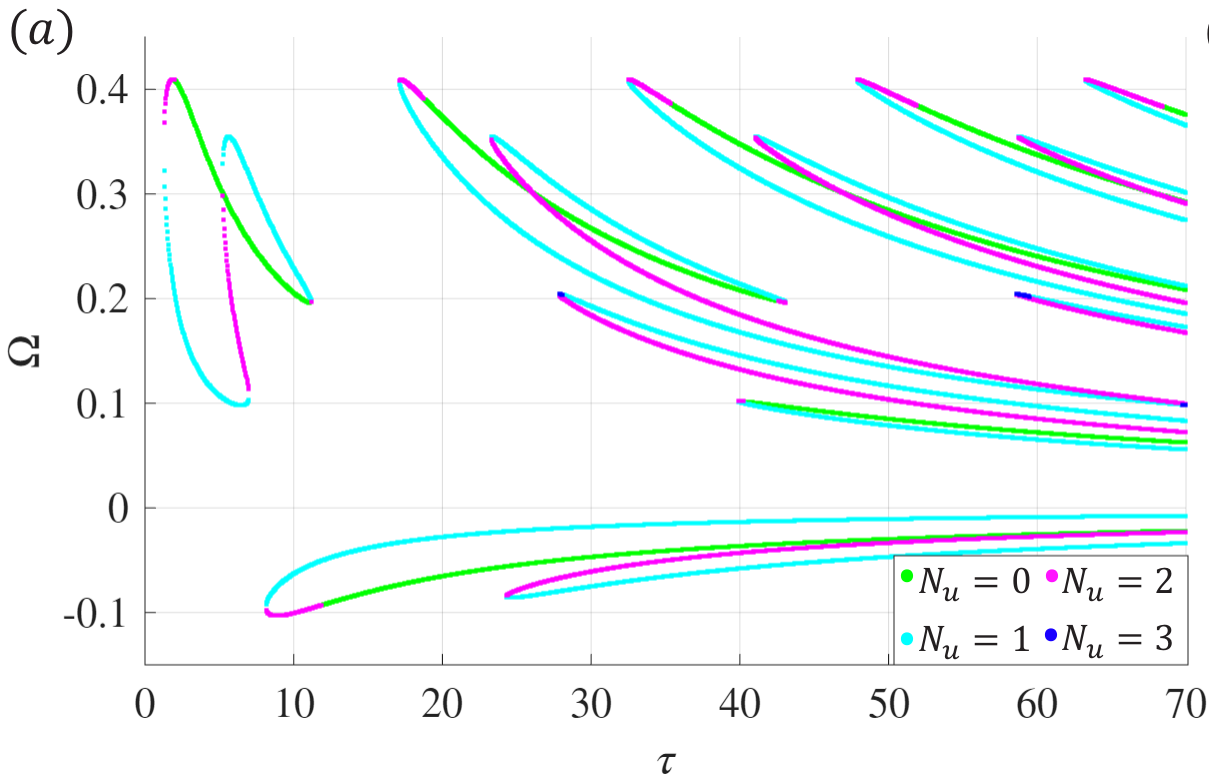}
\includegraphics[width=0.92\linewidth]{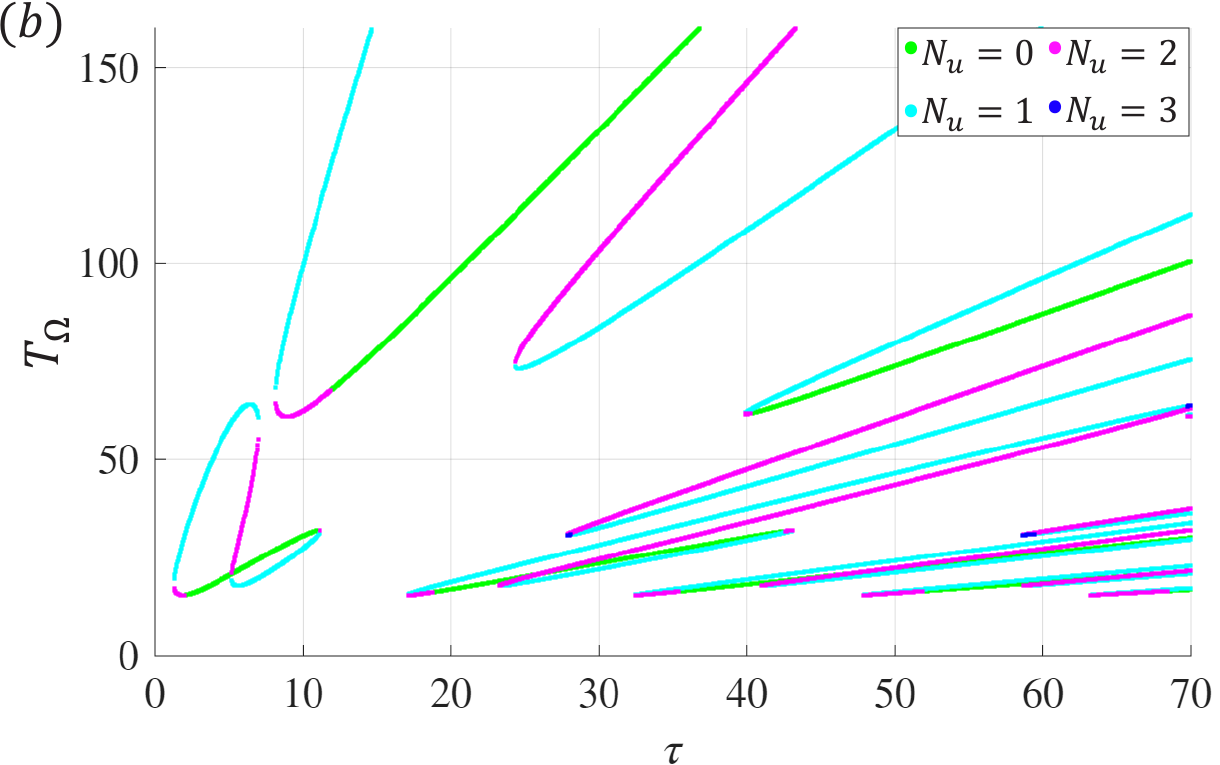}
\caption{Relative-equilibrium branches continued in the delay $\tau$ using the parametric representation \eqref{eq:tau_gamma}--\eqref{eq:T_gamma} and DDE-BifTool (for stability): (a) oscillation frequency $\Omega$ and (b) oscillation period $T_{\Omega}$. The color coding indicates $N_u$, the number of eigenvalues with positive real part (green is stable). The remaining parameters are $\omega_1=0.2$, $\omega_2=0.1$, $\alpha=\frac{\pi}{4}$, $a_1=0.5$, $a_2=0.4$, and $\epsilon=4\times10^{-4}$.}
\label{fig:RE_delay}
\end{figure}

System \eqref{eq:Fast_Omega} can be further reduced to a scalar equation for $\Omega$ and solved numerically, as it was done in the previous Sec.~\ref{subsec:CM_REs}. However, here we proceed differently: using Eq.~\eqref{eq:Fast_Omega}, we express $\Omega$ and $\tau$ explicitly in  parametric form $(\Omega(\gamma),\tau(\gamma))$. This allows us to obtain the analytical bifurcation diagram $\Omega,\tau$ (see Fig.~\ref{fig:RE_delay}). 

The parametric representation $(\Omega(\gamma),\tau(\gamma))$ can be done as follows. Denote $\gamma=\Omega\tau+\alpha$, then Eq.~\eqref{eq:Fast_Omega} can be rewritten as 
\begin{equation}
\label{eq:PQ}
\begin{aligned}
P(\gamma)=\cos\gamma\cos(2\theta)-\sin\gamma\sin(2\theta),\\
Q(\gamma)=\cos\gamma\sin(2\theta)-\sin\gamma\cos(2\theta),
\end{aligned}
\end{equation}
where 
\begin{align*}
    P(\gamma) =&\cos\gamma+\frac{2\left(\gamma-\alpha-\omega_{1}\tau\right)}{a_1\tau}\\
    Q(\gamma)=&\sin\gamma-\frac{2\left(\gamma-\alpha-\omega_{2}\tau\right)}{a_2\tau}.
\end{align*}
For $\cos(2\gamma)\ne0$, solving for $\cos(2\theta)$ and $\sin(2\theta)$, 
and using the identity $\cos^2(2\theta)+\sin^2(2\theta)=1$, we obtain a quadratic equation for the delay $\tau$:
\begin{align}
    A(\gamma)\tau^2+B(\gamma)\tau+C(\gamma)=0,
\label{eq:ABC_tau_quadratic}
\end{align}
where
\begin{align*}
A(\gamma)=&\left[\frac{1}{2}-\frac{\omega_1\cos\gamma}{a_1}+\frac{\omega_2\sin\gamma}{a_2}\right]^2
\\&+\left[\frac{\sin2\gamma}{2}-\frac{\omega_1\sin\gamma}{a_1}+\frac{\omega_2\cos\gamma}{a_2}\right]^2
-\frac{\cos^2 2\gamma}{4},
\end{align*}
\begin{align*}
B(\gamma)=&2(\gamma-\alpha)\Bigg[D_1(\gamma)\left(\frac{1}{2}-\frac{\omega_1\cos\gamma}{a_1}+\frac{\omega_2\sin\gamma}{a_2}\right)\notag\\&\quad
+D_2(\gamma)\left(\frac{\sin2\gamma}{2}-\frac{\omega_1\sin\gamma}{a_1}+\frac{\omega_2\cos\gamma}{a_2}\right)\Bigg],
\end{align*}
\begin{align*}
C(\gamma)=&(\gamma-\alpha)^2\left(\left(D_1(\gamma)\right)^2+\left(D_2(\gamma)\right)^2\right],
\end{align*}
\begin{align*}
D_1(\gamma)=&\frac{\cos\gamma}{a_1}-\frac{\sin\gamma}{a_2},
D_2(\gamma)=\frac{\sin\gamma}{a_1}-\frac{\cos\gamma}{a_2}.
\end{align*}
The nongeneric case $\cos(2\gamma)=0$ must instead be treated directly from Eq.~\eqref{eq:PQ}.
For $A(\gamma)\ne 0$ and $B(\gamma)^2-4A(\gamma)C(\gamma)\geq 0$, we have
\begin{align}
\tau_\pm(\gamma)=\frac{-B(\gamma)\pm
\sqrt{B(\gamma)^2-4A(\gamma)C(\gamma)}}
{2A(\gamma)}.
\label{eq:tau_gamma}
\end{align}

Together with the relations
\begin{align}
    \Omega(\gamma)
    =
    \frac{\gamma-\alpha}{\tau},
   \qquad
   T_{\Omega}(\gamma)=\frac{2\pi}{|\Omega(\gamma)|},
        \label{eq:T_gamma}
\end{align}
Eqs.~\eqref{eq:tau_gamma}--\eqref{eq:T_gamma} provide explicit parametric way of representing $\Omega$ and $T_\Omega$ as functions of $\tau$. 
If $A(\gamma)=0$ and $B(\gamma)\ne0$, Eq.~\eqref{eq:ABC_tau_quadratic} reduces to the linear relation $\tau=-C(\gamma)/B(\gamma)$; more degenerate cases must be checked directly from Eq.~\eqref{eq:ABC_tau_quadratic}. 

Figure~\ref{fig:RE_delay} shows the results of the application of Eqs.~\eqref{eq:tau_gamma}--\eqref{eq:T_gamma}: the relative-equilibrium branches of the slow-fast system \eqref{eq:phi1}--\eqref{eq:kappa2} as functions of the delay $\tau$, and their corresponding periods $T_{\Omega}$. 
Additionally, the stability of the relative equilibria are computed numerically using an extension of DDE-BifTool \cite{sieber2014ddebiftool} that accounts for phase-shift symmetry. 
The resulting branches show that, for a fixed $\tau$, multiple relative equilibria with distinct oscillation frequencies coexist. 
These coexisting relative equilibria can have different instability indices $N_u$, where $N_u$ denotes the number of characteristic roots with positive real parts.

\bibliography{references}

\end{document}